\RequirePackage{fix-cm}
\documentclass[smallextended]{svjour3}       
\smartqed  
\usepackage{graphicx}

\usepackage{amsfonts, amsmath, amssymb, latexsym, eepic}
\usepackage{subfigure}
\usepackage{paralist}
\usepackage{url}
\usepackage{color,ulem}
\usepackage{caption}

\newcommand{\be}[1]{\begin{equation}\label{#1}}
\newcommand{\ee}{\end{equation}}

\newcommand{\bsp}[1]{\begin{equation}\label{#1}\begin{split}}
\newcommand{\esplit}{\end{split}\end{equation}}

\newcommand{\dem}[1]{\textcolor{black}{#1}}

\begin{document}

\title{Post-buckling behaviour of a growing elastic rod}

\author{Axel A. Almet \and Helen M. Byrne
\and Philip K. Maini \and Derek E.~Moulton}

\institute{A.A. Almet \at 
		Wolfson Centre for Mathematical Biology, Mathematical Institute, University of Oxford, Oxford, UK
		\and
		H.M. Byrne \at 
		Wolfson Centre for Mathematical Biology, Mathematical Institute, University of Oxford, Oxford, UK
		\and
		P.K. Maini  \at 
		Wolfson Centre for Mathematical Biology, Mathematical Institute, University of Oxford, Oxford, UK
		\and
		D.E. Moulton  \at 
		Wolfson Centre for Mathematical Biology, Mathematical Institute, University of Oxford, Oxford, UK
 \email{moulton@maths.ox.ac.uk}
 }

\date{Received: date / Accepted: date}

\maketitle

\begin{abstract}
We consider mechanically-induced pattern formation within the framework of a growing, planar, elastic rod attached to an elastic foundation. Through a combination of weakly nonlinear analysis and numerical methods, we identify how the shape and type of buckling (super- or subcritical) depend on material parameters, and a complex phase-space of transition from super- to subcritical is uncovered. We then examine the effect of heterogeneity on buckling and post-buckling behaviour, in the context of a heterogeneous substrate adhesion, elastic stiffness, or growth. We show how the same functional form of heterogeneity in different properties is manifest in a vastly differing post-buckled shape. Finally, a fourth form of heterogeneity, an imperfect foundation, is incorporated and shown to have a more dramatic impact on the buckling instability, a difference that can be qualitatively understood via the weakly nonlinear analysis.

\end{abstract}

\section{Introduction}
Mechanically-induced pattern formation is a phenomenon prevalent in the morphogenesis of many biological structures, from airway wall remodelling \cite{Moulton2011}, to wrinkling of skin \cite{Efimenko2005}, to blades of grass \cite{Dervaux2009}. The prevailing feature in such systems is the deformation from a `trivial' base state to a more complex geometry, with buckling induced by mechanical stress. This feature is not unique to biological systems; indeed the same basic wrinkling pattern found in an elephant's skin can be induced by compressing a sheet of rubber. What is unique to the biological world is that such patterns tend to form without any external influence, rather the stress needed for mechanical instability is produced internally. \dem{Stress can be introduced internally via different mechanisms, including apical purse string contraction \cite{Sawyer:2010ku}, muscle contraction \cite{Dick:2018dy}, and uniform growth in a confined geometry.  A primary origin for stress, and the focus of this paper, is differential growth}, i.e. different parts of a tissue growing at different rates. Perhaps the simplest example is a tissue layer that grows relative to an underlying substrate to which it is adhered. The growth induces a compressive stress in the growing layer, and at some critical threshold the tissue buckles, exchanging compression energy for bending energy.  This situation underlies the formation of numerous biological patterns, including brain development---induced by the differential growth between the cortex and subcortex \cite{Budday2015}; intestinal crypt fission, in which epithelial tissue grows but is tethered to underlying tissue stroma \cite{Wong2002}; even seashell ornamentation, characterised by the adhesion of the growing mantle tissue to the rigid shell that it secretes \cite{Chirat2013}.   Aside from differential growth, inherent geometrical constraints can also induce buckling, for instance a row of cells growing uniformly but within a closed space will similarly develop compressive stress and ultimately buckle.

A mathematical description of growth-induced mechanical buckling can take a variety of forms. A number of discrete cell-based models have been devised, in which mechanical interactions between individual cells, coupled with cell growth or proliferation, can lead to deformation in the form of folds \cite{Drasdo2001}, invaginations \cite{Odell1981}, or protrusions \cite{Buske2012,Langlands2016}. This approach is more amenable to the inclusion of cell-level biological detail. Continuum modelling, while less amenable to this level of detail, allows one to utilise analytical tools for differential equations, which may improve insight and reveal parametric relationships that are more difficult to attain from discrete models.  Within 3D elasticity, growth is naturally incorporated via decomposition of the deformation gradient tensor into a growth tensor describing the local change of mass and an elastic tensor accounting for the elastic response \cite{Rodriguez1994}. However, beyond simple geometries such as the buckling of a sphere \cite{BenAmar2005}, a 3D description of buckling typically requires fully computational techniques such as finite element methods  \cite{Ambrosi2011}. In many cases, the geometry under consideration is well-suited for a reduced dimensional analysis. This is clearly true in the case of filaments.  Filaments by definition have one length scale much longer than the other two and hence are well suited to a 1D description. Kirchhoff theory for elastic rods has been applied to a diverse range of filamentary systems, such as DNA coiling \cite{Thompson2002}, neurite motility \cite{Recho2016}, plant tendril twisting \cite{Goriely1998}, and many more. A planar rod description may also be relevant even in an inherently 2D system: for instance when a sheet of tissue deforms approximately uniformly in a transverse direction, or when a cross-section of tissue deforms such as in the circumferential wrinkling of a tube \cite{moulton2011circumferential,Amar2013,Balbi2013}. 

Within a continuum formalism, a commonly-considered problem is a planar rod on an elastic foundation or substrate, typically with the ends fixed. The basic premise is that growth (axial elongation) of the rod generates compressive stress, creating buckling from a flat to a curved state, but this deformation is resisted by elastic tethering to a fixed substrate. This basic setup, or very similar, has been studied for many years in an engineering context (where `thermal expansion' typically plays the role of `growth'), dating back to classic works of Timoshenko \cite{Timoshenko1925} and Biot \cite{Biot1937}, and continues to find new interest and applications. 

It is only more recently that the relation to biological pattern formation has become clear and similar systems have been specialised for biological problems. An important aspect that distinguishes biological systems from the above examples is the various ways in which growth can occur; a key challenge here is connecting a continuum level description of growth to underlying cell-level processes. This has stimulated extensive mathematical modelling development and has created the need for a systematic framework. There are three such works of particular relevance for the present paper. In Moulton et al.\ \cite{Moulton2013}, the theory of Kirchhoff rods was extended by incorporating growth in a manner inspired by the morphoelastic decomposition. This is the framework upon which our analysis is built. Also of note are recent descriptions of buckling in the context of intestinal crypt formation, invaginations that are present throughout the intestines. Edwards and Chapman \cite{Edwards2007} applied a continuum mechanics approach to the formation of a single crypt. They modelled the crypt epithelium and its underlying tissue stroma as a beam upon a viscoelastic foundation. By performing a linear stability and eigenvalue analysis of buckling, they examined the effect of changes to proliferation, cell death, adhesion, or motility. Nelson et al. \cite{Nelson2011} complemented this analysis with a `bilayer' model representing an epithelial layer connected to a flexible substrate.  Nelson et al. also conducted an eigenvalue analysis similar to  Edwards and Chapman and combined this with a numerical analysis of the full, nonlinear model, demonstrating the influence of heterogeneity in both growth and bending stiffness on the resulting buckled crypt shape. Similar systems, but in 2D using plate theory, have also explored pattern formation due to growth instabilities. Hannezo et al. \cite{Hannezo2011} characterised transitions from crypt-like to herringbone and labryinth patterns; Nelson et al. extended their 1D models in \cite{nelson2013buckling} and found that crypt patterning could be most strongly controlled through heterogeneity in growth.

Our objective in this paper is to use the morphoelastic rod framework of Moulton et al.\ \cite{Moulton2013} to extend the results of  \cite{Edwards2007,Nelson2011} and analyse unexplored features that are of general relevance. A focal point for our analysis is the behaviour of the system beyond the initial buckling. In an engineering context, buckling will typically signify failure, and so the threshold value to induce buckling may be the most relevant quantity. For pattern formation in biology, on the other hand, the shape evolution well beyond the initial instability is often critical to the final pattern (and its biological functionality), and hence analysis only of the onset of instability is insufficient. 

Also of relevance in many biological systems is understanding the role of heterogeneity in mechanical pattern formation. Heterogeneity can arise in three main forms: growth\footnote{Here we refer to non-uniform growth along the axis of the buckling tissue, not the differential growth that is assumed to occur between layers.}, the mechanical properties of the rod, or substrate adhesion. Here it is important to distinguish between growth and remodelling. Growth refers to an increase (or decrease) in mass, i.e. a change in size of a tissue layer without any change in its material properties. Remodelling, on the other hand, refers to a change in material properties without any change in mass, e.g. due to fibre reorientation or cell differentiation. In a growing tissue, both of these processes occur and will commonly occur non-uniformly.  The crypt, for instance, is not a layer of uniform cells, but rather consists of a clear proliferative hierarchy of cells with varying rates of division as one moves up the crypt axis \cite{Wright1984}.  Heterogeneity in adhesion may occur due to non-uniform changes in the substrate layer, or in a biological context due to changes in the cells, or may occur due to buckling itself, for instance due to viscoelastic effects.

Both of these features---large deformation beyond buckling and heterogeneity---pose significant mathematical challenges.  To capture post-buckling behaviour requires analysis of a nonlinear system of equations, as opposed to the linear stability analysis that can be used to detect buckling. Furthermore, including heterogeneity complicates the use of many analytical tools, either rendering the system analytically intractable or complicating attempts to unfold bifurcations. Here, rather than rely fully on computational techniques, our approach is to analyse post-buckling behaviour and the effect of heterogeneity through a combination of a weakly nonlinear analysis and numerical solution.  This approach yields a broad understanding of the role of heterogeneities in growth, material properties, and adhesion, and reveals features of post-buckling pattern formation not described in previous analyses.

We consider a 1D model system of a growing planar rod on an elastic foundation, serving both as an extension of the classical setup and as an abstracted framework for several of the aforementioned biological systems. The rod is subject to growth in the axial direction and clamped boundary conditions, which drive buckling at a critical growth. The goal of this paper is to understand the factors driving the onset of buckling and the post-buckling behaviour. In particular, we investigate how the buckling and post-buckling behaviour changes in the presence of spatial heterogeneity in material properties, obtaining explicit relations for how the pitchfork bifurcation that arises is impacted by heterogeneity, and exploring the shape evolution in the nonlinear post-buckled regime.

The remainder of this paper is structured as follows. In Section 2, we outline the general theory for Kirchoff rods and incorporation of growth, as developed in \cite{Moulton2013}. Then, in Section 3 we summarise the linear stability analysis before extending to a weakly nonlinear analysis. The results of the weakly nonlinear analysis and numerical analysis of the full nonlinear model are presented in Section 4, first in a homogeneous setting, then with the addition of different material heterogeneities. Finally, we close by discussing the implications of our results and directions for future model extensions.

\section{Model setup}
\begin{figure}[t!]
\centering
\captionsetup{width = 0.85\textwidth}
\includegraphics[scale=0.17]{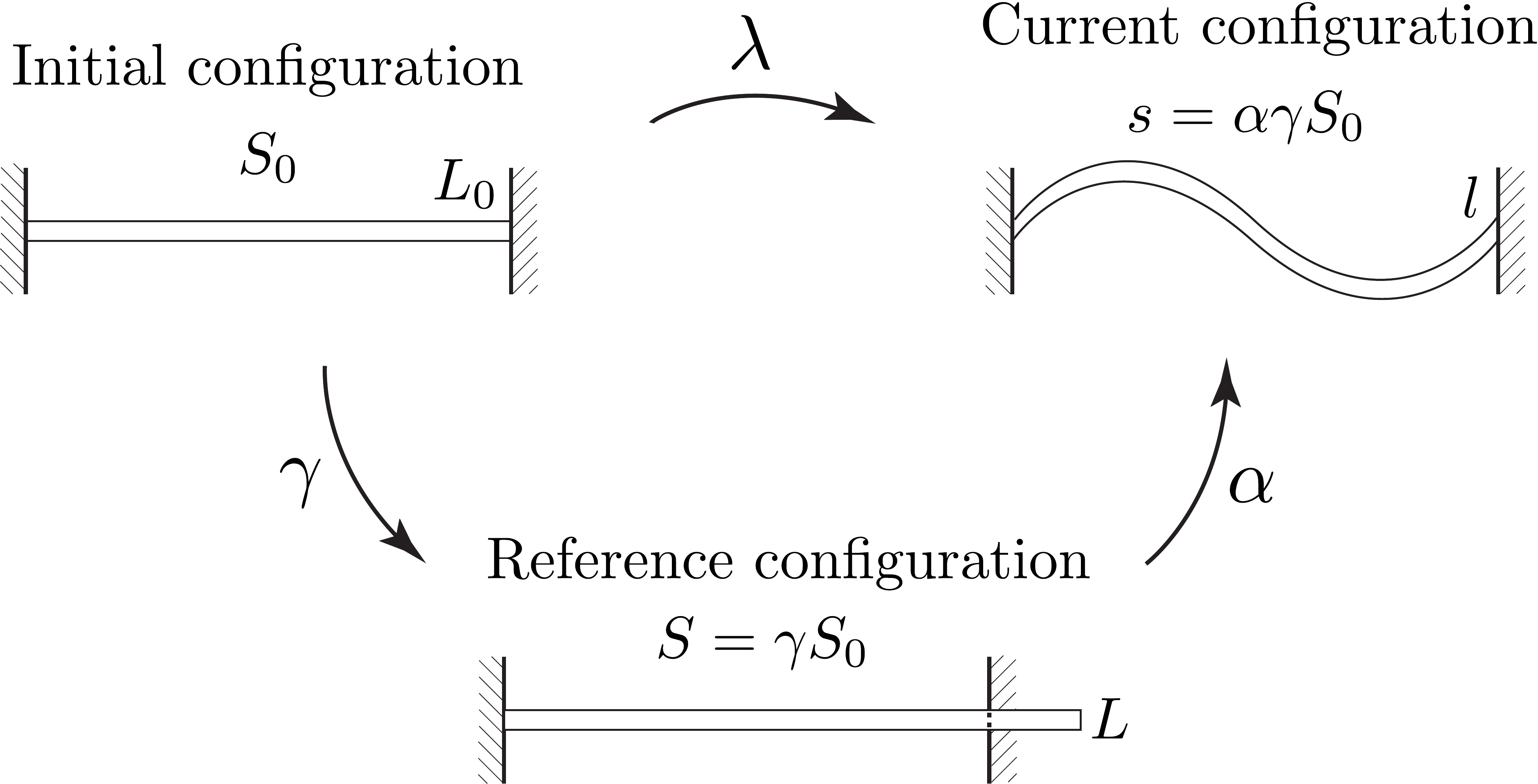}
\caption{\textbf{The different morphoelastic rod configurations.} A rod that is initially confined to a (finite) interval grows in a virtual, unstressed reference configuration, before being mapped to the current configuration, where it is subject to boundary conditions and loads. The rod is parametrised by a different arc length in each configuration. The respective rod lengths have been indicated. The parameters $\gamma$, $\alpha$ and $\lambda$ denote the growth, elastic, and total stretches, respectively.}
\label{fig:planarmorphoroddiagram}
\end{figure}

As a model system to investigate post-buckling in growing slender structures, we consider in this paper an extensible and unshearable planar rod in quasi-static mechanical equilibrium. The rod is constrained geometrically by clamped-clamped boundary conditions, and is also adhered to an elastic (Winkler) foundation. Growth of the rod is modelled under the morphoelastic rod framework of \cite{Moulton2013}. In this framework, one identifies three distinct configurations: an initial reference configuration parametrised by the arc length $S_0$; a grown virtual configuration, parameterised by $S$ and referred to as the `reference' configuration; and the current configuration, parametrised by $s$. This framework is summarised in Figure \ref{fig:planarmorphoroddiagram}. In the initial, reference, and current configurations, the total rod lengths are $L_0$, $L$, and $l$, respectively.  The rod arc length is assumed to evolve through an axial growth process, described by the growth stretch\footnote{\dem{Here we follow standard terminology \cite{Goriely:2016tc} in referring to $\gamma$ as the growth 'stretch', the rationale being that growth acts to 'stretch' the arclength by adding new material. That is, the word `stretch' does not refer to a `stretching' of old material, but rather an increase in reference arclength by the addition of new material.}} $\gamma(S_0) = \partial S/\partial S_0$, followed by an elastic response, encapsulated by the elastic stretch $\alpha$. The total stretch of the rod $\lambda$ from initial to current configuration is then given by 
\begin{align}
\lambda = \alpha\gamma \Leftrightarrow \frac{\partial s}{\partial S_0} = \frac{\partial s}{\partial S}\frac{\partial S}{\partial S_0}.\label{eq:morphoelasticityassumption}
\end{align}
This is the 1D analogue of the multiplicative decomposition of the deformation gradient tensor employed in 3D morphoelasticity \cite{Rodriguez1994,BenAmar2005,Moulton2013}. 

Let the rod's centerline be given by $(x,y)$, and let $\theta$ denote the angle between the tangent vector, $\mathbf{d}_3 = \cos\theta\mathbf{e}_x + \sin\theta\mathbf{e}_y$, and the $x$-axis. This is expressed by 
\be{}
\frac{\partial x}{\partial S_0} =\alpha\gamma\cos\theta, \qquad \frac{\partial y}{\partial S_0} = \alpha\gamma\sin\theta,\label{eq:geomeqnsdimsimplified}\\
\ee
(The scale factor $\alpha\gamma$ accounts for the fact that we parametrise the system by the initial arclength parameter $S_0$, as opposed to the current arclength parameter $s$.)
Defining the resultant force and moment in the rod by $\mathbf{n} = F\mathbf{e}_x + G\mathbf{e}_y$ and $\mathbf{m} = m\mathbf{e}_z$, the balance of linear and angular momentum give
\begin{align}
&\frac{\partial F}{\partial S_0} + \gamma f = 0,\qquad \frac{\partial G}{\partial S_0} + \gamma g = 0,\label{eq:forceeqnsdimsimplified}\\
& \frac{\partial m}{\partial S_0} + \alpha\gamma(G\cos\theta - F\sin\theta) = 0.\label{eq:momenteqnsdimsimplified}
\end{align}
Here $\mathbf{f} = f\mathbf{e}_x + g\mathbf{e}_y$ is the external body force, which is assumed to be solely due to the underlying foundation. 

To this system we add a constitutive equation relating moment to curvature:
\begin{align}
m = \frac{EI}{\gamma}\frac{\partial\theta}{\partial S_0},\label{eq:simplifiedconstitutivemoment}
\end{align}
The parameter $E$ is the Young's modulus of the rod, and is taken for now to be constant, and $I$ denotes the (second) moment of inertia. For an extensible rod, we take the axial stress to be related to the elastic stretch $\alpha$ via a linear constitutive relation
\begin{align}
F\cos\theta + G\sin\theta = EA(\alpha - 1),\label{eq:constitutivetensionrefconfig}
\end{align}
where $A$ is the cross-sectional area of the rod. 

The foundation is assumed to be a linearly elastic medium occupying an interval along the $x$-axis, as is the rod. Initially the foundation is a distance $y_0$ from the rod centreline and the rod is glued to the $x$-axis; that is, a point $(S_0, 0)$ along the $x$-axis is attached to a point $(S_0, y_0)$ on the rod. We can set $y_0 = 0$ without loss of generality. No remodelling takes place, so these two points are still connected in the reference configuration, and are now at $(S/\gamma, 0)$ and $(x, y)$ respectively. Therefore the body force acting on the rod (as a force per initial length) is
\begin{align}
f\mathbf{e}_x + g\mathbf{e}_y = -\frac{Ek}{\gamma}\left[(x-S_0)\mathbf{e}_x+y\mathbf{e}_y\right],\label{eq:foundationconstitutivelawdim}
\end{align}
where we assume the foundation stiffness is proportional to that of the rod, with the dimensionless positive parameter $k$ comparing the stiffness of the foundation to that of the rod. The factor of $1/\gamma$ indicates that the body force is parametrised with respect to the initial configuration, and that no remodelling occurs after growth.

The system is closed with the clamped boundary conditions at $S_0=0$ and $L_0$:
\begin{align}
x(0) = 0, \quad x(L_0) = L_0, \qquad y(0) = y(L_0) = 0, \qquad \theta(0) = \theta(L_0) = 0.\label{eq:dimBCs}
\end{align}
\subsection{Non-dimensionalisation}
Next, we non-dimensionalise the system using the standard Kirchhoff scaling \cite{Coleman1993,Goriely1997,Goriely1997a} and circumflexes to denote non-dimensional quantities:
\begin{align}
\{S_0, x, y \} =(A/I)^{1/2}\left\{\widehat{S}_0,\widehat{x}, \widehat{y}\right\}, \quad \{F,G\} = EA\left\{\widehat{F}, \widehat{G}\right\}, \quad m = E(AI)^{1/2}\widehat{m}.\label{eq:nondimensionalisation}
\end{align}
Dropping the circumflexes of independent and dependent variables for notational convenience, Equations \eqref{eq:geomeqnsdimsimplified}--\eqref{eq:momenteqnsdimsimplified} simplify to
\begin{align}
&\frac{\partial x}{\partial S_0} =\alpha\gamma\cos\theta, \qquad \frac{\partial y}{\partial S_0} = \alpha\gamma\sin\theta,\label{eq:geomeqnsnondimsimplified}\\
&\frac{\partial F}{\partial S_0} = \widehat{k}(x - S_0), \qquad \frac{\partial G}{\partial S_0} = \widehat{k}y,\label{eq:forceeqnsnondimsimplified}\\
&\frac{\partial\theta}{\partial S_0} = \gamma m, \qquad \frac{\partial m}{\partial S_0} + \alpha\gamma(G\cos\theta - F\sin\theta) = 0.\label{eq:momenteqnsnondimsimplified}
\end{align}
The constitutive law for extensibility now reads
\begin{align}
&F\cos\theta + G\sin\theta = \alpha - 1.\label{eq:constitutivetensionnondim}
\end{align}
The dimensionless boundary conditions are
\begin{align}
x(0) = 0, \quad x(\widehat{L}_0) = \widehat{L}_0, \qquad y(0) = y(\widehat{L}_0) =  0, \qquad \theta(0) =  \theta(\widehat{L}_0) = 0,\label{eq:nondimBCs}
\end{align}
where the remaining (non-dimensional) parameters are
\begin{align}
\widehat{L}_0 = L_0(A/I)^{1/2}, \qquad \widehat{k} = \frac{kI}{A^2}.\label{eq:nondimparams}
\end{align}
Note that the non-dimensional rod length $\widehat{L}_0$ depends on the ratio of two length characteristics of the rod: its initial total length $L_0$ and the thickness, characterised by $(I/A)^{1/2}$. For instance, for a rod with circular cross-section of radius $r$, we have $I=\pi r^4/4$, $A=\pi r^2$, and hence $\widehat{L}_0=2L_0/r\gg1$.
\section{Stability Analysis}
In this section, we present the analytical tools that we will use to investigate the buckling and post-buckling behaviour of the morphoelastic rod. We first adapt and summarise the linear stability analysis from \cite{Moulton2013}, used to calculate the growth bifurcation value, $\gamma^*$, before unfolding the bifurcation with a weakly nonlinear analysis.

\subsection{Linear stability analysis}
\label{sec:linearstabilityanalysis}
We first determine the critical growth stretch $\gamma^*$ and corresponding buckling mode using a linear stability analysis. The calculations in this section are also present in Moulton et al.\ \cite{Moulton2013}, but are summarised here for completeness and to motivate the weakly nonlinear analysis. Inspecting the system \eqref{eq:geomeqnsnondimsimplified}--\eqref{eq:nondimBCs}, for all $\gamma>1$ there exists a base solution corresponding to a straight, compressed rod; that is, with $\theta \equiv 0$ and the total stretch \dem{ $\lambda = \alpha\gamma=1$ (implying an elastic compression $\alpha=1/\gamma$)}:
\begin{align}
x^{(0)} = S_0, \qquad y^{(0)} = 0 \qquad F^{(0)} = \frac{1-\gamma}{\gamma}, \qquad G^{(0)} = \theta^{(0)} = m^{(0)} = 0.\label{eq:trivialsolution}
\end{align}
Expanding each variable about the base solution so that, for example, $x = x^{(0)} + \delta x^{(1)} + O(\delta^2)$, where $\delta$ is an arbitrary small parameter, and considering $O(\delta)$ terms, leads to the linearised system
\begin{align}
&x^{(1)'}-\gamma F^{(1)} = 0, \qquad y^{(1)'} - \theta^{(1)} = 0,\label{eq:geomeqnslinearised}\\
&F^{(1)'} - \widehat{k}x^{(1)} = 0, \qquad G^{(1)'} - \widehat{k}y^{(1)} = 0,\label{eq:forceeqnslinearised}\\
&\theta^{(1)'} - \gamma m^{(1)} = 0, \qquad m^{(1)'} + \frac{(\gamma-1)}{\gamma}\theta^{(1)} + G^{(1)} = 0,\label{eq:momenteqnslinearised}
\end{align}
where $' = \partial/\partial S_0$. We observe that $\{x^{(1)}, F^{(1)}\}$ are decoupled from $\{y^{(1)}, G^{(1)}, m^{(1)}, \theta^{(1)}\}$, and that \eqref{eq:geomeqnslinearised}-\eqref{eq:momenteqnslinearised} can be expressed as two ordinary differential equations for $x^{(1)}$ and $y^{(1)}$,
\begin{align}
&Lx^{(1)} := x^{(1)''} - b^2x^{(1)} = 0,\label{eq:xdirectioneqnlinearised}\\
&My^{(1)} := y^{(1)''''} + 2ay^{(1)''} + b^2y^{(1)} = 0,\label{eq:ydirectioneqnlinearised}
\end{align}
where the coefficients $a$ and $b$ are defined by
\begin{align}
a = \frac{\gamma - 1}{2}, \qquad b = \left(\widehat{k}\gamma\right)^{\frac{1}{2}}.\label{eq:aandb}
\end{align}
Solving Equation \eqref{eq:xdirectioneqnlinearised} subject to $x^{(1)}(0)=x^{(1)}(\widehat L_0)=0$ leads to the trivial solution $x^{(1)} \equiv 0$ (and subsequently $F^{(1)} \equiv 0$). Turning to Equation \eqref{eq:ydirectioneqnlinearised}, our linearised boundary conditions are 
\begin{align}
y^{(1)} = y^{(1)'} = 0 \qquad\mbox{ at }\qquad S_0 = 0,\ \widehat L_0.\label{eq:linearisedbcs}
\end{align}
Seeking solutions of the form $y^{(1)} \sim e^{i\omega S_0}$ (valid on an infinite domain) leads to the following oscillation frequencies
\begin{align}
\omega_{\pm}^2 = a \pm \left(a^2 - b^2\right)^{\frac{1}{2}}.\label{eq:omega1and2}
\end{align}
We remark that in order for non-damped oscillations to exist over the (finite) domain, we require $a \ge b$. Applying the four boundary conditions for $y^{(1)}$, specified by Equation \eqref{eq:linearisedbcs}, yields the solution
\begin{align}
y^{(1)} = C_1\left[\cos(\omega_+S_0) - \cos(\omega_-S_0) + C_2\sin(\omega_+S_0) + C_3\sin(\omega_-S_0)\right],\label{eq:y1solution}
\end{align}
where the constants $C_2$ and $C_3$ are given by
\begin{align}
C_2 = \frac{\omega_-(\cos(\widehat{L}_0\omega_-) - \cos(\widehat{L}_0\omega_+))}{\omega_-\sin(\widehat{L}_0\omega_+) - \omega_+\sin(\widehat{L}_0\omega_-)}, \qquad C_3 =  \frac{\omega_+(\cos(\widehat{L}_0\omega_+) - \cos(\widehat{L}_0\omega_-))}{\omega_-\sin(\widehat{L}_0\omega_+) - \omega_+\sin(\widehat{L}_0\omega_-)},\label{eq:C2andC3}
\end{align}
and the critical growth value $\gamma^*$ must satisfy the relation
\begin{align}
a\sin(\widehat{L}_0\omega_+)\sin(\widehat{L}_0\omega_-) + b\cos(\widehat{L}_0\omega_+)\cos(\widehat{L}_0\omega_-) - b = 0.\label{eq:gammarooteqn}
\end{align}
If \eqref{eq:gammarooteqn} is satisfied, then the buckled solution is given by \eqref{eq:y1solution}, but with arbitrary constant $C_1$.  For later convenience, we thus define the function $\hat{y}$ to be the determined part of this function, i.e.

\begin{equation}
\widehat{y}:=C_1^{-1}y^{(1)}.\label{eq:yhatsolution}
\end{equation}
The smallest value of $\gamma > 1 $ that satisfies \eqref{eq:gammarooteqn} occurs when $a = b$ (and, hence, $\omega_1 = \omega_2$), giving
\begin{align}
\gamma^*_{\mathrm{inf}} = 1 + 2\widehat{k} + 2\left(\widehat{k} + \widehat{k}^2\right)^{\frac{1}{2}}\label{eq:gammazero},
\end{align}
where $\widehat{k} = kI/A^2$. Although it appears that $\gamma^*_\mathrm{inf}$ does not vary with any length scale, this is not entirely true. For example, for a rod with circular cross-section of radius $r$, then from \eqref{eq:nondimparams}, $\widehat{k} \propto k$. However, for a rectangular cross-section with height $h$ and width $w$,  $\widehat{k} \propto kh/w$. Additionally, this value of $\gamma$ only leads to oscillations if the rod length $\widehat{L}_0$ is infinite, and results in the trivial solution over a finite domain. Therefore the critical growth value $\gamma^*$ is the first value of $\gamma > \gamma^*_{\mathrm{inf}}$ that solves \eqref{eq:gammarooteqn}.

\subsection{Weakly nonlinear analysis}
\label{sec:weaklynonlinearanalysis}
Having summarised the above results from \cite{Moulton2013} that are key for us, we now carry out a weakly nonlinear analysis. For a given root of \eqref{eq:gammarooteqn}, the buckled solution is only determined to within the arbitrary constant $C_1$. In order to understand the behaviour of the buckled rod as it continues to grow, it is necessary to determine how the buckling amplitude $C_1$ depends on $\gamma$, which we accomplish through a weakly nonlinear analysis. We unfold the bifurcation by introducing the ansatz
\begin{align}
\gamma = \gamma^* + \varepsilon\gamma^{(1)}\label{eq:growthunfolding},
\end{align}
where $\varepsilon$ is a fixed small parameter (whereas $\delta$ is small, but arbitrary) and $\gamma^{(1)} = O(1)$ is a control parameter describing the proximity to the growth bifurcation point $\gamma^*$. Substituting \eqref{eq:growthunfolding} into \eqref{eq:momenteqnsnondimsimplified}, re-expanding $\theta$ and $m$ and retaining higher-order terms in $\theta$ reveals that the nonlinearities will be balanced by growth if $\varepsilon = O(\delta^2)$. Setting $\varepsilon = \delta^2$ and re-expanding our variables about the trivial solution \eqref{eq:trivialsolution} as a power series in $\delta$ leads to the following system of differential equations for each order $O(\delta^n)$, $n \ge 1$,
\begin{align}
&x^{(n)'}-\gamma^*F^{(n)} = h_{x^{(n)}}, \qquad y^{(n)'} - \theta^{(n)} = h_{y^{(n)}},\label{eq:geomexpandedn}\\
&F^{(n)'} - \widehat{k}x^{(n)} = h_{F^{(n)}}, \qquad G^{(n)'} - \widehat{k}y^{(n)} = h_{G^{(n)}},\label{eq:forceexpandedn}\\
&\theta^{(n)'} - \gamma^*m^{(n)} = h_{\theta^{(n)}}, \qquad m^{(n)'} + \frac{(\gamma^*-1)}{\gamma^*}\theta^{(n)} + G^{(n)} = h_{m^{(n)}}.\label{eq:momentgeneraln}
\end{align}
Here, the functions $h_{x^{(n)}}, h_{y^{(n)}}, h_{F^{(n)}}, h_{G^{(n)}}, h_{\theta^{(n)}}$, and  $h_{m^{(n)}}$ denote  inhomogeneities due to lower order terms. As was the case in Section \ref{sec:linearstabilityanalysis}, the system decouples into two linear operators acting on $x^{(n)}$ and $y^{(n)}$,
\begin{align}
&Lx^{(n)} = h'_{x^{(n)}} + \gamma^*h_{F^{(n)}} =: H_{x^{(n)}},\label{eq:xdirectioneqngeneraln}\\
&My^{(n)}  = h'''_{y^{(n)}} + 2ah'_{y^{(n)}} + h''_{\theta^{(n)}} + \gamma^*\left(h'_{m^{(n)}} - h_{G^{(n)}}\right) =: H_{y^{(n)}}\label{eq:ydirectioneqngeneraln}.
\end{align}
For general $n \ge 1$, the boundary conditions for $y^{(n)}$ are now given by
\begin{align}
y^{(n)} = 0, \qquad y^{(n)'} = h_{y^{(n)}} \qquad\mbox{ at }\qquad S_0 = 0,\ \widehat{L}_0.\label{eq:yBCsgeneraln}
\end{align}
Observe that the homogeneous problems, \eqref{eq:xdirectioneqnlinearised} and \eqref{eq:ydirectioneqnlinearised}, along with the boundary conditions---$x^{(1)}(0) = x^{(1)}(\widehat{L}_0) = 0$ and  \eqref{eq:linearisedbcs}, respectively---are self-adjoint.   \dem{When $n = 1$, we recover the linearised system described by Equations \eqref{eq:geomeqnslinearised}--\eqref{eq:momenteqnslinearised}, i.e. $H_{x^{(1)}} = H_{y^{(1)}} \equiv 0$.
At $n = 2$ we find that $H_{y^{(2)}} \equiv 0$, giving us no further information on the buckling amplitude $C_1$; however, $H_{x^{(2)}}$ is nonlinear in $y^{(1)}$ and thus a non-trivial solution for $x^{(2)}$ exists,} as $x^{(1)}$ is trivial and the Fredholm Alternative Theorem is immediately satisfied. Hence, we must consider $O(\delta^3)$ terms to obtain the amplitude equation for $C_1$. This leads us to consider the inhomogeneities $H_{y^{(3)}}$, which can be expressed solely in terms of $x^{(2)}$ and $y^{(1)}$ and their derivatives. Then, by the Fredholm Alternative Theorem, a solution for $y^{(3)}$ exists if and only if the following solvability condition is satisfied:
\begin{align}
\int^{\widehat{L}_0}_0 \left(My^{(3)}\right)\widehat{y}\;dS_0 = \int^{\widehat{L}_0}_0 \left(H_{y^{(3)}}\right)\widehat{y}\;dS_0 = 0,\label{eq:y3solvability}
\end{align}
where $\hat{y}$ solves the homogeneous problem, defined by \eqref{eq:yhatsolution}. The only unknown quantity in \eqref{eq:y3solvability} is the coefficient $C_1$; hence
this solvability condition for $y^{(3)}$ provides a relation between the buckling amplitude $C_1$ and the distance from the critical buckling growth parameter, expressed by $\gamma^{(1)}$. Simplifying this condition (see Appendix A) leads us to deduce that
\begin{equation}
C_1\left(K_1C_1^2 + K_2\gamma^{(1)}\right) = 0.\label{eq:bucklingamplitudeeqn}
\end{equation}
The constants $K_1$, $K_2$ are tedious to compute analytically, but nevertheless only depend only on the material parameters $\hat k$ and $\widehat{L}_0$, and through them the critical growth $\gamma^*$. It can also be shown (Appendix A) that $K_2>0$ for all parameter choices.  Equation \eqref{eq:bucklingamplitudeeqn} shows that the system exhibits a pitchfork bifurcation, a known property of similar systems \cite{Nelson2011,Hutchinson1970}. The three branches of the pitchfork are given by 
\begin{equation}
C_1 = 0, \qquad  C_1^2 =- \frac{K_2}{K_1}\gamma^{(1)}.\label{eq:bucklingamplitudes}
\end{equation}
This shows that the bifurcation will be supercritical if $K_1<0$, and subcritical if $K_1>0$. In the next section, we explore the dependence of  $K_1$ on $\widehat{k}$ and $\widehat{L}_0$ and its effect on the buckling and post-buckling behaviour.

\section{Buckling and post-buckling behaviour}
\label{sec:results}
Having established a relationship for the post-buckling amplitude, we now explore the effect of material parameters and heterogeneity on the buckling and post-buckling behaviour. First we examine the form of bifurcation in a homogeneous setting, effectively by analysing how the critical buckling growth $\gamma^*$, the buckling mode, and the pitchfork constants $K_1$ and $K_2$ vary with the two free parameters in the non-dimensional system, $\widehat L_0$ and $\hat k$.  We then adapt the weakly nonlinear analysis to incorporate heterogeneity and investigate the impact of non-uniformity in foundation stiffness, rod stiffness, and growth. In each case, we complement the analytical work by solving the full nonlinear system \eqref{eq:geomeqnsnondimsimplified}--\eqref{eq:nondimBCs}, using the numerical package \texttt{AUTO-07p} \cite{Doedel2007}. \texttt{AUTO-07p} uses pseudo-arclength continuation to trace solution families and solves the system with an adaptive polynomial collocation method.

\subsection{Effect of length and foundation stiffness}
\label{sec:wnaverificationvaryingk}
\begin{figure}[t!]
\centering
\captionsetup{width=\textwidth}
\includegraphics[width=\textwidth]{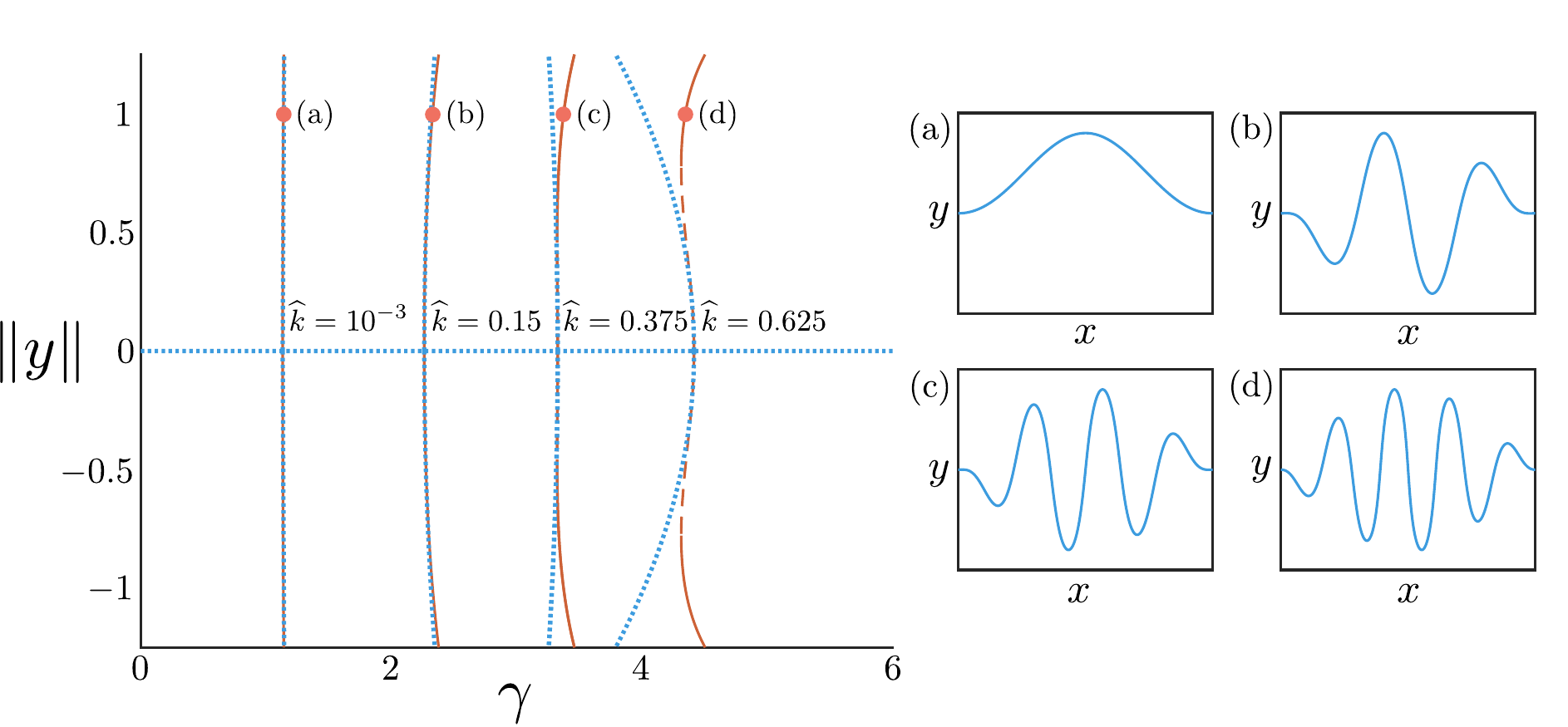}
\caption{\textbf{Bifurcation diagram for varying foundation stiffness.} Bifurcation diagram plotting $\pm\|y\|=\pm\text{max}_{S_0}\vert y(S_0)\vert$ against $\gamma$ from weakly nonlinear analysis (blue, dotted lines) and numerical continuation (red) of the full model \eqref{eq:geomeqnsnondimsimplified}--\eqref{eq:nondimBCs}; stable solutions are marked with solid lines, while dashed lines represent unstable solutions. Each non-trivial branch denotes a continuation in $\gamma$ for a fixed value of $\widehat{k}$. The buckled shape $(x(S_0), y(S_0))$ is plotted at the point $\|y\| = 1$ (orange dot) for (a) $\widehat{k} = 10^{-3}$, (b) $\widehat{k} = 0.15$, (c) $ \widehat{k} = 0.375 $, and (d) $\widehat{k} = 0.625$. The dimensionless rod length is fixed to be $\widehat{L}_0 = 20$ for all cases.}
\label{fig:wnaverification}
\end{figure}

\begin{figure}[t!]
\centering
\captionsetup{width=\textwidth}
\includegraphics[width=\textwidth]{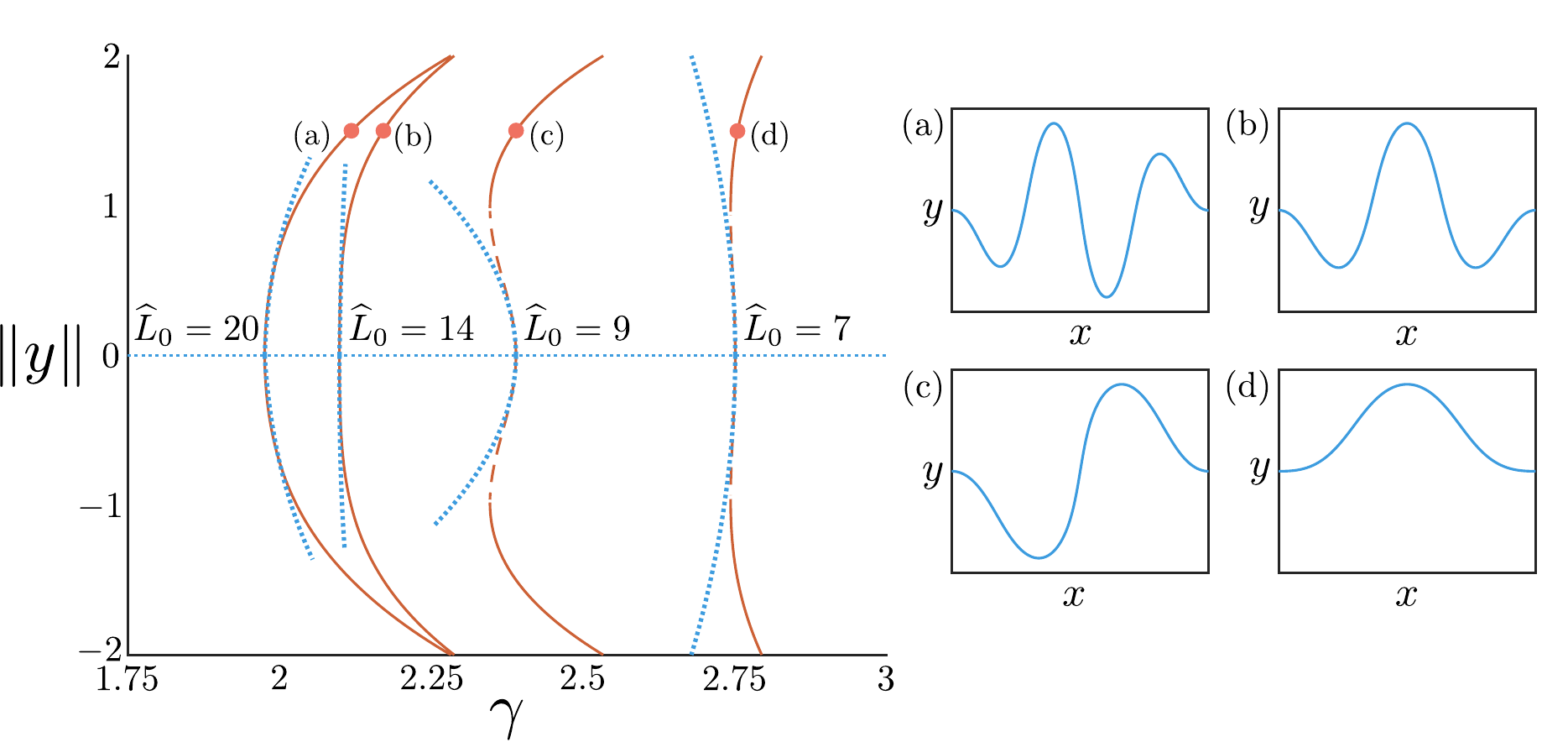}
\caption{\textbf{Bifurcation diagram for varying rod length.} Bifurcation diagram plotting $\pm\| y\|=\pm\text{max}_{S_0}\vert y(S_0)\vert$ against $\gamma$ from weakly nonlinear analysis (blue, dotted lines) and numerical continuation (red) of the full model \eqref{eq:geomeqnsnondimsimplified}--\eqref{eq:nondimBCs}; stable and unstable solutions are marked with solid and dashed lines respectively. The buckled shape $(x(S_0), y(S_0))$ is plotted at the point $\|y\| = 1.5$ (orange dot) for (a) $\widehat{L}_0 = 20$, (b) $\widehat{L}_0 = 14$, (c) $ \widehat{L}_0 = 9$, and (d) $\widehat{L}_0 = 7$. The dimensionless foundation stiffness $\widehat{k}$ is set to $\widehat{k} = 0.1$ for all cases.}
\label{fig:wnaverificationvaryinglength}
\end{figure}

In Figures \ref{fig:wnaverification}--\ref{fig:wnaverificationvaryinglength}, we plot bifurcation diagrams for varying values of the dimensionless foundation stiffness $\widehat{k}$ and the dimensionless rod length $\widehat{L}_0$, respectively, for increasing growth.  The horizontal axis in each case is the growth parameter $\gamma$, and the vertical axis plots the non-trivial branches, $\pm\|y\|:=\pm\text{max}_{S_0}\vert y(S_0)\vert$, which is closely related to the value of the constant $C_1$ but more representative of the post-buckling amplitude. The solid lines are determined from the weakly nonlinear analysis, while the dashed lines are numerical results. We also plot the buckled shape $(x(S_0), y(S_0))$ at the specified points for each branch.

As $\widehat{k}$ increases, the buckling occurs for increased mode number, reflecting the energy trade-off that as the foundation stiffness is increased, a large amplitude is penalised more by a high foundation energy, and hence higher bending energy is sacrificed to have a lower amplitude. 
An increased value of $\widehat{k}$ also leads to an increase in $\gamma^*$, which shows that the foundation is serving to stabilise the rod against buckling. This can again be understood in terms of an energy trade-off, but in this case it is the compressive energy in the grown but unbuckled state that is being sacrificed.  It is important to note that this feature could not occur in an inextensible rod, for which the trivial state does not exist for any $\gamma>1$.
 
Considering length $\widehat{L}_0$ leads to similar changes in both critical buckling growth and mode number. However, while the buckling mode increases for increasing length, the critical buckling growth {\it decreases}.  Recalling the scaling $\widehat{L}_0 = L_0(A/I)^{1/2}$, this reflects the notion that a short or thick rod can endure more growth before buckling, and will buckle at lower mode.
We note also that as $\widehat{L}_0\to\infty$, $\gamma^*\to \gamma^*_{\mathrm{inf}}$, the critical growth value for buckling on an infinite domain \eqref{eq:gammazero}.

Perhaps most notable is the transition from supercritical to subcritical bifurcation evident in both diagrams.  We find that subcritical bifurcations occur for large enough $\widehat{k}$ or small enough $\widehat{L}_0$.  We find through numerical continuation that the subcritical branches then fold back, a feature not captured by the weakly nonlinear analysis at order $O(\delta^3)$. A linear stability analysis (see Appendix B) confirms that the portion of the subcritical branch before folding back is unstable, while the portion after the fold-back is stable. (As would be expected, the curved branches are stable in the supercritical case.)  This implies that a subcritical bifurcation signifies a discontinuous jump from the trivial flat state to the finite amplitude stable branch, as well as the presence of a hysteresis loop if $\gamma$ is subsequently decreased.	

\subsection{Locating the pitchfork transition}
\begin{figure}[t!]
\hspace*{-0.5cm}
\centering
\captionsetup{width=\textwidth}
\includegraphics[width=0.8\textwidth]{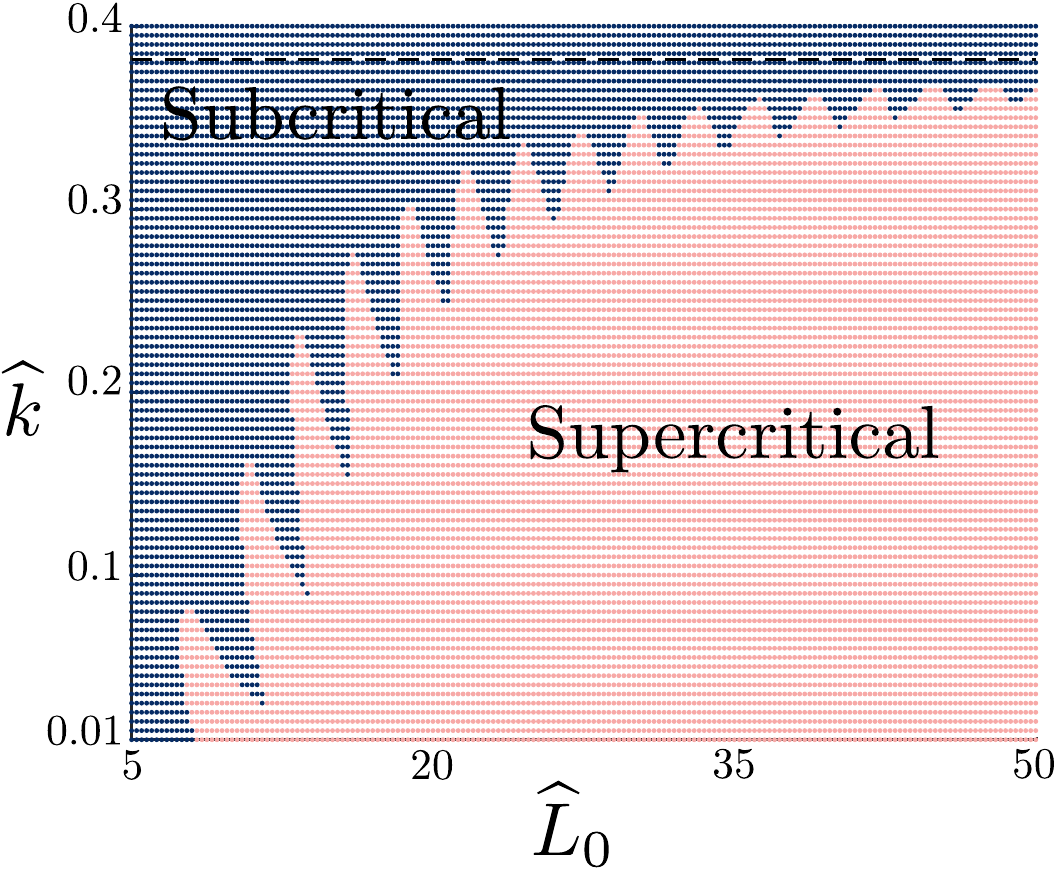}
\caption{\textbf{Phase diagram of pitchfork bifurcations.} The regions have been determined by computing the sign of $K_1$ at each point on a discretised grid of $\widehat{k}$ and $\widehat{L}_0$. Subcritical pitchfork bifurcations ($K_1 > 0$) have been labelled with dark blue crosses, while supercritical pitchfork bifurcations are labelled with an orange dot ($K_1 < 0$). The dashed line corresponds to the transition value of $\widehat{k}$ in the infinite-length case, $\widehat{k} \approx 0.38196$ (see Appendix \ref{sec:pitchforkinfinitedomain}).}
\label{fig:pitchforkchange}
\end{figure}

For given parameters $\widehat{k}$ and $\widehat{L}_0$, the weakly nonlinear analysis enables us to determine the type of pitchfork bifurcation simply by computing the sign of $K_1$. Figure \ref{fig:pitchforkchange} shows the regions in $\widehat{k}$-$\widehat{L}_0$ space where supercritical and subcritical pitchforks occur.  Note that despite having an explicit expression for $K_1$, the actual computation of its value was done numerically as it requires root finding for the eigenvalue $\gamma^*$. Hence, to produce Figure \ref{fig:pitchforkchange} we computed $K_1$ over a discrete grid in the $\widehat{k}$-$\widehat{L}_0$ plane. The transition boundary was then verified at several points through numerical path following in \texttt{AUTO-07p}.

Unexpectedly, we do not find a simple monotonic transition boundary, as seen in similar studies \cite{Hutchinson2013}, but rather an intricate pattern with an oscillatory structure.   This structure implies that multiple transitions between super- and subcritical buckling can occur for a fixed $\widehat{k}$ and varying $\widehat{L}_0$ (or vice versa); that is, simply increasing the length of the rod monotonically can create repeated transitions between super- and subcritical bifurcation. For an infinite rod, the transition can be computed as $\widehat{k} \approx 0.38196$ (see Appendix \ref{sec:pitchforkinfinitedomain}); this point is included as a dashed, horizontal line in Figure \ref{fig:pitchforkchange}, and it appears that as $\widehat{L}_0\to\infty$, the oscillations dampen and the transition boundary approaches this constant value. By contrast, as $\widehat{L}_0$ decreases, the oscillation amplitude increases, although the slenderness assumption of the rod breaks down as $\widehat{L}_0\sim O(1)$. 

\dem{This intricate structure, which to our knowledge has not been reported before, has interesting potential implications. The defining characteristic of the subcritical regime is a discontinuous bifurcation: a small change in growth beyond the critical value leads to a potentially large jump in amplitude; while the bifurcation is smooth in the supercritical regime. Hence, the fact that the transition boundary oscillates in the parameter space implies a sort of non-robustness to the instability.  The effects of subcritical bifurcations have been studied in the biological contexts of biochemical Turing patterns \cite{BrenaMedina:2014ij}, epidemics \cite{Ruan:2003vx}, and even neuroscience \cite{Laing:2002we}. The phenomenon seems to be less-well studied in mechanical models of morphogenesis, despite clear analogies with engineering structures where subcritical buckling is well-documented, e.g. \cite{Magnusson:2001vj}. This may be in part due to the difficulty in observing the actual instability event in biological morphogenesis, hence classifying the form of bifurcation is not straightforward. Subcritical bifurcation in a mechanical context has however been observed in a model of the buckling of a lipid bilayer vesicle between two plates \cite{preston2008buckling}. It remains an interesting open question whether or not the transition region in Fig. \ref{fig:pitchforkchange} could be physically realised. Of course the model system presented here is highly idealised, and the complexities of the structure may either not exist in a real system or be detectable within experimental error; nevertheless the framework could in principle be tailored to a particular biological setting to explore the form of bifurcation in greater detail.}   

\subsection{Parameter heterogeneity}
\label{sec:parameterheterogeneity}
Thus far, we have assumed spatial homogeneity in model parameters. However, in many biological systems, heterogeneities are inherent in the system. This raises the question of how a given heterogeneity is manifest in the buckled pattern. To explore this, we initially consider three distinct forms of heterogeneity in: the foundation stiffness, the rod stiffness, and the growth. Each heterogeneity has a clear biological interpretation. For example, in the intestinal crypts, these heterogeneities would correspond to: the different types of extracellular matrix secreted by the cells comprising the underlying tissue stroma (foundation heterogeneity), the mechanical properties of epithelial cells in the crypt (stiffness heterogeneity), and the variations in the proliferative capacity of these cells (growth heterogeneity). In order to understand the effect of each type of spatial heterogeneity, we examine heterogeneity for each parameter in isolation. 

With parameter heterogeneity, it becomes increasingly difficult to obtain analytically tractable results with a weakly nonlinear analysis, especially if the amplitude of the heterogeneity is pronounced. Nevertheless, when the parameters are close to homogeneous, we can extend the weakly nonlinear analysis and, in particular, ask how the heterogeneity impacts the pitchfork bifurcations observed in the homogeneous case.  We complement this analysis with numerical solutions of the full system defined by Equations \eqref{eq:geomeqnsnondimsimplified}--\eqref{eq:nondimBCs}. For computational convenience, heterogeneity is incorporated via a sequence of numerical continuations in the growth and heterogeneity parameters.

We model heterogeneity as a spatial deviation from a baseline homogeneous state. In general, for an arbitrary parameter $\mu$, we consider
\begin{align}
\mu(S_0) = \mu_0 + \widehat\epsilon\mu_0\xi(S_0).\label{eq:arbitraryheterogeneity}
\end{align}
Here, the constant $\mu_0$ corresponds to the baseline homogeneous value, and the function $\xi(S_0)$ captures the spatial variation, modulated by the amplitude factor $\widehat\epsilon$ and constrained only by the requirement that $\mu\geq0$. 

With heterogeneity introduced, the weakly nonlinear analysis can be viewed as a two-parameter unfolding, both in the distance from the critical buckling growth, via $\gamma = \gamma^* + \varepsilon\gamma^{(1)}$, and in the distance from homogeneity, characterised by $\widehat{\epsilon}$.   We are thus faced with balancing three small parameters: $\varepsilon$,  $\widehat\epsilon$, and the order of the expanded variables, which we denoted by $\delta$, e.g. as in $y = \delta y^{(1)} + O(\delta^2)$. In the homogeneous case the correct balance is given by $\epsilon=\delta^2$. With $\widehat\epsilon > 0$, numerous balances could be sought, and a full analysis of the two-parameter unfolding is beyond the scope of this paper. Our approach involves starting from homogeneity, and increasing the order of $\widehat\epsilon$ to see when and how it first impacts on buckling.  Hence, we again take $\epsilon=\delta^2$, and consider $\widehat\epsilon=\delta^\beta\eta$, with $\beta>1$ and $\eta$ an $O(1)$ control parameter.

The three cases we wish to consider for heterogeneity are:

\begin{itemize}
\item {\bf Foundation heterogeneity}, for which $\mu=\widehat{k}$;
\item {\bf Rod stiffness heterogeneity}, for which $\mu=E$, the Young's modulus\footnote{In this case heterogeneity is incorporated prior to non-dimensionalization, and scaling proceeds using the baseline value. Note also that we do not vary the Young's modulus $E$ present in the definition of the foundation force \eqref{eq:foundationconstitutivelawdim}, so that we may distinguish the material properties in the foundation from material properties of the rod itself. };
\item {\bf Growth heterogeneity}, for which $\mu=\gamma$.
\end{itemize}

Perturbing each of these parameters via \eqref{eq:arbitraryheterogeneity} has a similar effect on the weakly nonlinear analysis. In each case, it is easy to show that for $\beta>2$, the heterogeneity does not affect the weakly nonlinear analysis up to $O(\delta^3)$, and therefore does not affect the buckling amplitude $C_1$. Consequently, the bifurcation relation \eqref{eq:bucklingamplitudes} is unaffected. When $\beta=2$, the heterogeneity first has an impact (up to $O(\delta^3)$) and, hence, it is for this case that we adapt the analysis. At $O(\delta^3)$, the corrective term $y^{(3)}$ now satisfies
\begin{align}
My^{(3)} = H^\mathrm{old}_{y^{(3)}} + \eta H^\mathrm{new}_{y^{(3)}}.\label{eq:newy3eqn}
\end{align}
The first term on the right hand side, $H^\mathrm{old}_{y^{(3)}}$, corresponds to the inhomogeneities in the homogeneous case, while $H^\mathrm{new}_{y^{(3)}}$ describes the effects of the heterogeneity \eqref{eq:arbitraryheterogeneity}. 

For each parameter considered, evaluating the solvability condition \eqref{eq:y3solvability} leads to a new equation for $C_1$:
\begin{align}
&C_1\left(K_1C_1^2 + K_2\gamma^{(1)} + K_3\eta \right) = 0.\label{eq:bucklingamplitudeeqnvariablefoundation}
\end{align}
The constants $K_1$ and $K_2$ are identical those in Equation \eqref{eq:bucklingamplitudeeqn} (see Appendix \ref{sec:bucklingamplitudec1}). The heterogeneity is fully encapsulated in the term $K_3$, defined straightforwardly by
\begin{align}
K_3 = C_1^{-1}\int^{\widehat{L}_0}_0 H^\mathrm{new}_{y^{(3)}}\ \widehat{y}\ dS_0. \label{eq:K3defn}
\end{align}

The heterogeneous model, hence, undergoes a translated pitchfork bifurcation, where the branches are given by
\begin{align}
&C_1 = 0, \qquad C_1^2 = -\frac{K_2}{K_1}\gamma^{(1)} -\frac{K_3}{K_1}\eta.\label{eq:bucklingamplitudesvariablefoundation}
\end{align}
Observe that since $\xi$ appears in $H^\mathrm{new}_{y^{(3)}}$ only, and hence in $K_3$ only, it does not affect the type of pitchfork that occurs, but merely translates it. That is, setting $C_1 = 0$ in the non-trivial branch gives $\gamma^{(1)} = -(K_3/K_2)\eta$; recalling \eqref{eq:growthunfolding}, the critical growth $\gamma^*$ is now shifted to
\begin{align}
\gamma^* = \gamma_0^* - \delta^2\frac{K_3}{K_2}\eta,\label{eq:criticalgammavariablefoundation}
\end{align}
where $\gamma_0^*$ is the critical growth stretch for the rod in a homogeneous setting, as determined from the linear stability analysis in Section \ref{sec:linearstabilityanalysis}. Hence, we see that the material heterogeneity \eqref{eq:arbitraryheterogeneity} results in an $\mathcal{O}(\delta^2)$ shift in $\gamma^*$.  Since $K_2>0$, the direction and degree of the shift is determined by $K_3$. Generally, $K_3$ provides the `metric' for whether the heterogeneity has a net effect of strengthening or weakening the effect of the material parameter.

\begin{figure}[t!]
\centering
\captionsetup{width=\textwidth}
\includegraphics[width=\textwidth]{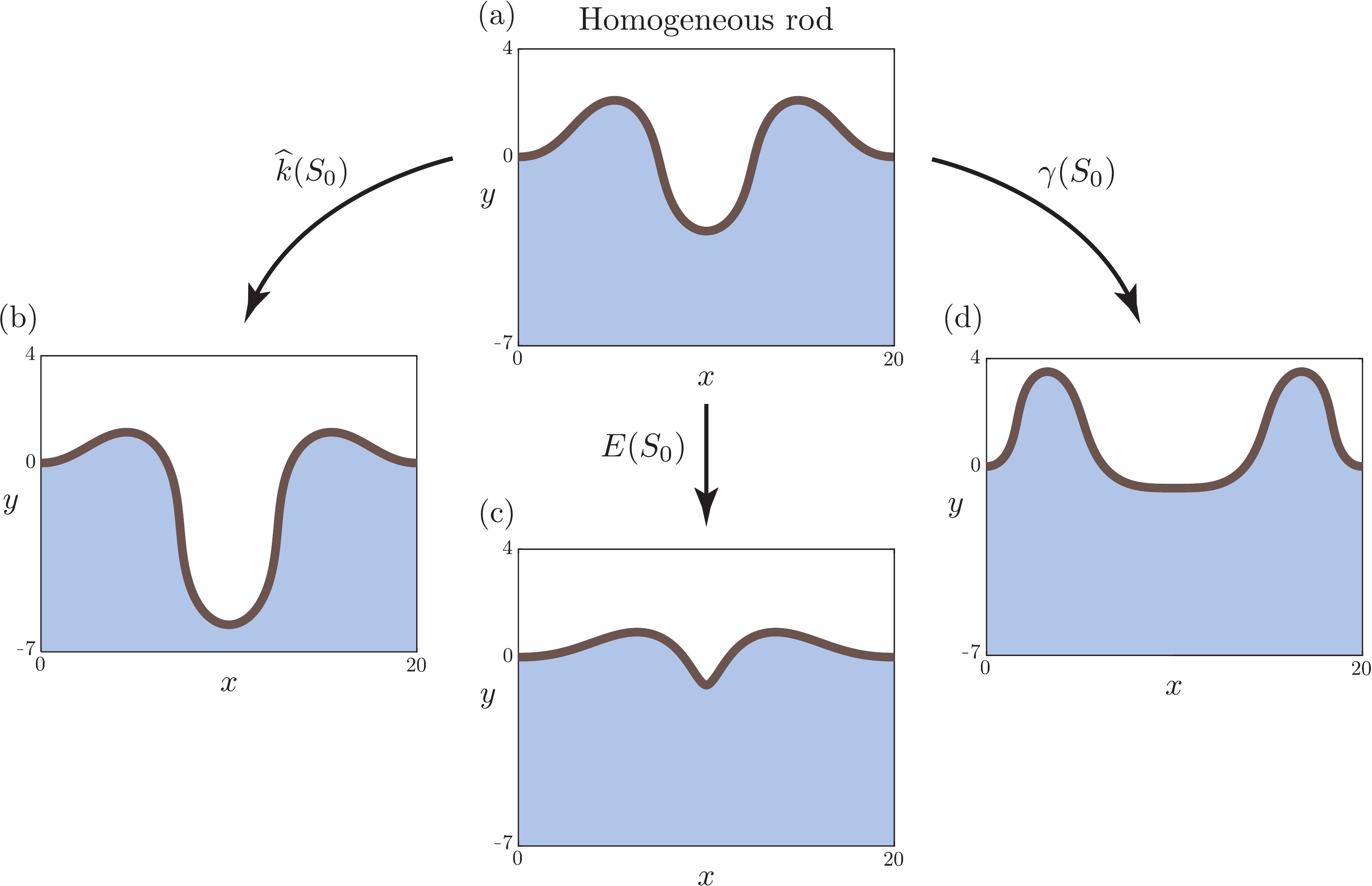}
\caption{\textbf{The effect of heterogeneity on the post-buckled shape $(x(S_0), y(S_0))$ and the underlying foundation}. The baseline foundation stiffness $\widehat{k}_0$ and rod length $\widehat{L}_0$ have been set to $\widehat{k} = 0.04$ and $\widehat{L}_0 = 20$, respectively. (a) The growth stretch $\gamma$ ($\gamma_0$ for $\gamma(S_0)$) has been continued until $\|y \| = 2.75$ for the homogeneous case. The heterogeneity function $\xi(S_0)$ has been set to $\xi(S_0) = \cos\left(\frac{2\pi S_0}{\widehat{L}_0}\right)$ for each of (b) foundation stiffness heterogeneity $\widehat{k}(S_0)$, (c) rod stiffness heterogeneity $E(S_0)$, and (d) growth heterogeneity  $\gamma(S_0)$. The heterogeneity amplitude $\widehat{\epsilon}$ has been continued to $\widehat{\epsilon} = 0.9$ from the homogeneous state ($\widehat{\epsilon} = 0$). }
\label{fig:heterogeneitysummarised}
\end{figure}

In order to investigate greater amplitudes of heterogeneity and the post-buckling shape evolution, we perform numerical continuation on the full model. As an illustrative example, we apply the same form of heterogeneity for each of the three parameters: $\xi(S_0) = \cos(2\pi S_0/\widehat{L}_0)$ and $\widehat{\epsilon} = 0.9$, characterising a significant decrease in the middle region and increase in the outer regions. Figure \ref{fig:heterogeneitysummarised} depicts the resultant rod shapes. As evident in Figure \ref{fig:heterogeneitysummarised}, the heterogeneity has a markedly different effect for each material parameter. 

\paragraph{Foundation stiffness heterogeneity} Here, the modified foundation is softer in the middle and stiffer near the endpoints, causing a significant increase in amplitude in the middle of the rod, where the resistance to deformation is weaker. This phenomenon can be generally understood and quantified by applying the weakly nonlinear analysis. Note from Equation \eqref{eq:ydirectioneqnlinearised} that the operator $M$ on the left hand side of Equation \eqref{eq:newy3eqn} is the linearised (beam) equation, i.e. the vertical force balance for an extensible rod upon a foundation. Consequently, the term $H_{y^{(3)}}^\mathrm{new}$ captures additional forces due to the imposed heterogeneity. In the case $\widehat{k}(S_0) = \widehat{k}_0(1 + \delta^2\eta\xi(S_0))$, at $O(\delta^3)$ this term takes the particularly simple and instructive form:
\begin{align}
H_{y^{(3)}}^\mathrm{new} = -\widehat{k}_0\gamma^*\xi\ y^{(1)}.\label{eq:Hy3newvariablefoundation}
\end{align}
The heterogeneity thus acts as an amplifying force where $\xi(S_0) < 0$, and a resistive force where $\xi(S_0) >0$. This is apparent in Figure \ref{fig:heterogeneitysummarised}(b): the magnitude of $y^{(1)}$ is largest in the middle, with $\xi(S_0) < 0$, reducing the effects of $H^\mathrm{old}_{y^{(3)}}$ and leading to an increase in amplitude.

\paragraph{Rod stiffness heterogeneity} In the case of rod stiffness, the dominant trend is compression in the middle region, leading to a significant decrease in amplitude and arclength, and formation of a near cusp-like point, reflecting  the reduced energy cost of both bending and stretching in the middle region.  We have also examined the competing energies within the system: bending versus stretching versus foundation (defined in Appendix \ref{sec:energy}). We find that both the bending and foundation energy are reduced as $\widehat{\epsilon}$ increases, despite the cusp-like formation, while the stretching energy increases (see  Figure \ref{fig:youngsheterogeneityenergyplot}). The total energy remains roughly constant through most of this tradeoff, but eventually, at large values of $\widehat{\epsilon}$, the stretching penalty outweighs the benefit to the bending and foundation energies and a sharp rise in the total energy occurs for $\widehat{\epsilon}\gtrsim0.7$.

\paragraph{Growth heterogeneity} In the case of growth heterogeneity, note that for the form of heterogeneity considered, $\xi(S_0) = \cos(2\pi S_0/\widehat{L}_0)$, the \emph{net growth}, defined by

\begin{align}
\overline{\gamma} = \frac{1}{\widehat{L}_0}\int^{\widehat{L}_0}_0\gamma(S_0)dS_0, \label{eq:netgrowth}
\end{align} 
is unchanged from the homogeneous case, $\gamma(S_0) = \gamma_0$. Thus, for varying $\widehat{\epsilon}$, there is no change in net growth, merely a redistribution of material from the middle region to the sides. Accordingly, Figure \ref{fig:heterogeneitysummarised}(d) shows a significant change in shape: the middle region flattens while the left and right regions, with increased material, show an increase in amplitude and curvature. The loss of material from the middle also has the effect of increasing the elastic stretch $\alpha$, resulting in a transition from compression to tension. An intuitive explanation for this can be seen by examining the flat solution \eqref{eq:trivialsolution}: a growth stretch of $\gamma < 1$ implies that the horizontal force $F^{(0)} > 0$, i.e. the rod is in a state of tension. For growth heterogeneity, we observe behaviour in energy that is qualitatively opposite from rod stiffness heterogeneity: as heterogeneity is increased, both bending and foundation energies increase, while stretching and total energies decrease. The flattening of the middle region caused by loss of material to the edges reduces the compressive energy locally, while the redistribution of material to the sides leads to a net increase in bending and foundation energy (Fig. \ref{fig:growthheterogeneityenergyplot}).

\dem{It is worth comparing these results to similar studies. In Nelson et al.\ \cite{Nelson2011}, bending stiffness heterogeneity and growth heterogeneity were considered. In the case of growth heterogeneity, Nelson et al.\ concluded that net growth affects the post-buckling behaviour more than heterogeneity, whereas we have found a significant change in morphology due to growth heterogeneity, even with no change in net growth. This discrepancy may be partially due again to the inextensibility assumption present in \cite{Nelson2011}. More likely though, the behaviour may be attributable to viscous relaxation. Nelson et al.\ have modelled the foundation as viscoelastic springs, thus incorporating a stress relaxation not present in our model. Indeed, they presented an example (see Fig. 11 of \cite{Nelson2011}) in which a change in morphology does initially occur due to growth heterogeneities, but the difference is then lost once stresses are allowed to relax. Some form of viscous relaxation is almost certainly present in development of the colorectal crypt, and incorporating such effects in our framework is the subject of ongoing work.}
\subsubsection{The role of extensibility.}

\begin{figure}[t!]
\centering
\captionsetup{width=\textwidth}
\includegraphics[width = 1.025\textwidth]{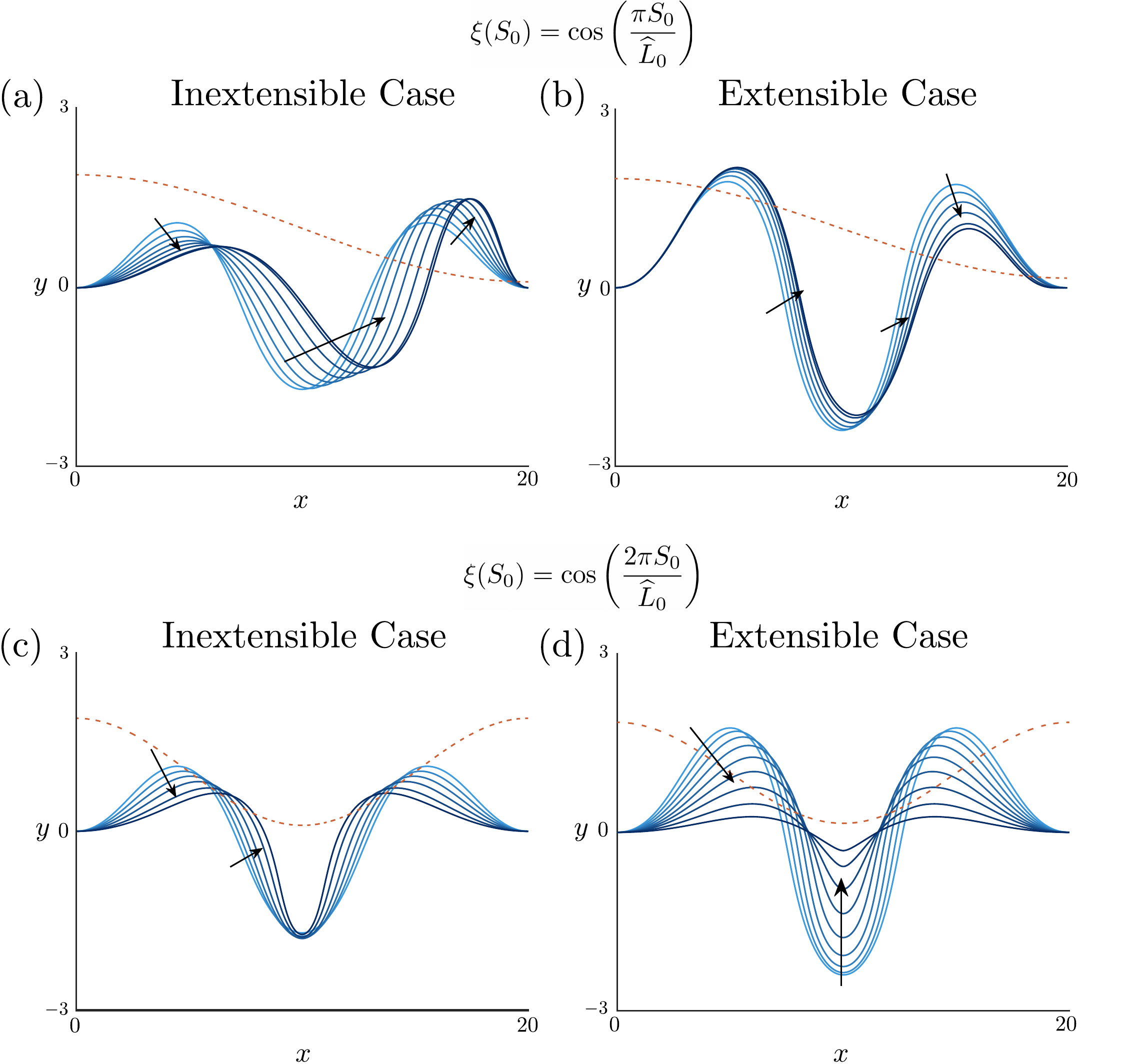}
\caption{\textbf{The effect of extensibility on rod shape.} The specified heterogeneities are (a)--(b) $\xi(S_0) = \cos\left(\frac{\pi S_0}{\widehat{L}_0}\right)$ and (c)--(d) $\xi(S_0) = \cos\left(\frac{2\pi S_0}{\widehat{L}_0}\right)$. The dimensionless foundation stiffness and rod length have been set to $\widehat{k} = 0.04$ and $\widehat{L}_0 = 20$, respectively. The growth parameter $\gamma$ was set to $\gamma = 1.1$ and $\gamma = 1.8$ for the inextensible and extensible cases respectively. Continuation in $\widehat{\epsilon}$ is over the interval $\widehat{\epsilon} \in [0, 0.85]$. The resulting forms of $E(S_0)$ (brown, dashed line) have also been plotted. Arrows in the plots indicate the evolution of the rod shape (blue, sold lines) in the increasing direction of the continuation parameter, while darker blue lines correspond to higher values of $\widehat{\epsilon}$. When rod stiffness is asymmetric, competition in curvature causes the inextensible rod to redistribute its material more so than the extensible rod. In the symmetric case, extensibility leads to compression at the locations of maximal curvature, which is not seen for the inextensible rod. }
\label{fig:youngsheterogeneitycomparison}
\end{figure}

It is important to note that many of the above trends are reliant on the assumption of rod extensibility. In an inextensible model, axial compression is not permitted, as the arclength is fixed, which is equivalent to the geometric constraint $\alpha \equiv 1$.
For explicit comparison, we consider the same stiffness heterogeneity in an inextensible rod.  Consequently, only bending is affected by the heterogeneity \eqref{eq:arbitraryheterogeneity}. In Figure \ref{fig:youngsheterogeneitycomparison}, we compare the shape evolution with increasing $\widehat{\epsilon}$ in both inextensible and extensible models, for $\xi(S_0) = \cos(\pi S_0/\widehat{L}_0)$ and $\xi(S_0) = \cos(2\pi S_0/\widehat{L}_0)$. For the inextensible models, we take an equivalent foundation stiffness, $\widehat{k} = 0.04$, but set $\gamma = 1.1$ to obtain a similar initial amplitude. In an inextensible rod, the arclength is fixed and thus the response to heterogeneity is to alter the shape towards aligning points of minimal and maximal curvature with material points of maximal and minimal stiffness, respectively. Hence in Figure \ref{fig:youngsheterogeneitycomparison}(a)  the inextensible rod shifts to have maximal amplitude on the soft region on the right side, whereas the extensible rod (Fig. \ref{fig:youngsheterogeneitycomparison}(b)) compresses on the right side, thus producing a completely different morphology with minimal amplitude. In Figure \ref{fig:youngsheterogeneitycomparison}(c), the inextensibility leads to a localisation of curvature in the soft middle region, as opposed to the strong compression in the extensible case, shown in Figure \ref{fig:youngsheterogeneitycomparison}(d). These simulations illustrate the dramatic effect that extensibility can have on shape morphology and the response to material heterogeneity.  

\dem{We note that Nelson et al. \cite{Nelson2011} also considered bending stiffness heterogeneity, finding that localised regions of softened bending stiffness leads to a localisation of buckling. This result is similar to Figure \ref{fig:youngsheterogeneitycomparison} for the inextensible case, where the rod shape shifts towards points of softened rod stiffness.}

\subsubsection{Foundation imperfection}
The heterogeneities we have considered thus far, while having significant impact on the post-buckling shape evolution, have had a relatively minor effect on the bifurcation itself, only serving to translate the pitchfork, and by modest amounts. This is in contrast to typical results in a Koiter imperfection sensitivity analysis \cite{Hutchinson1970,Amazigo:1970gs}, in which material imperfections may shift the bifurcation to occur at significantly reduced loads and in an imperfect fashion (a `broken pitchfork'). Here, the small change in bifurcation can be understood within the framework of our model by considering the form of heterogeneity imposed.  As derived in section \ref{sec:weaklynonlinearanalysis}, the base (homogeneous) equation for the pitchfork bifurcation is

$$K_1C_1^3 + K_2C_1 \gamma^{(1)} = 0.$$
Perturbations to the system in the form of heterogeneities have the potential to change this to
$$K_1C_1^3 + K_2C_1 \gamma^{(1)} + K_3C_1 + K_4 = 0.$$
The cases we have examined lead to $K_4=0$ and $K_3\neq0$, which merely translates the pitchfork (as $C_1=0$ is still a solution branch). Breaking the pitchfork would require $K_4\neq0$. The reason that the additional term obtained has a factor of $C_1$ is that we have only considered {\it multiplicative} heterogeneity, i.e. we have imposed heterogeneity in terms that multiply dependent system variables: foundation stiffness $k$ multiplies $x$ and $y$ in the force balance, stiffness $E$ multiplies $\theta'$ as well as $\alpha$, and growth $\gamma$ appears in the system multiplicatively in multiple places (as evident in Equations \eqref{eq:geomeqnsdimsimplified}-\eqref{eq:momenteqnsdimsimplified}). Thus, at the relevant order in an asymptotic expansion, a perturbation to these parameters always appears multiplicatively with the base solution $y_1=C_1\hat y$, and thus the additional term in the solvability condition that provides the bifurcation condition is of the form $K_3C_1$.  

\begin{figure}[t!]
\centering
\hspace*{0.25cm}
\captionsetup{width=\textwidth}
\includegraphics[width=0.85\textwidth]{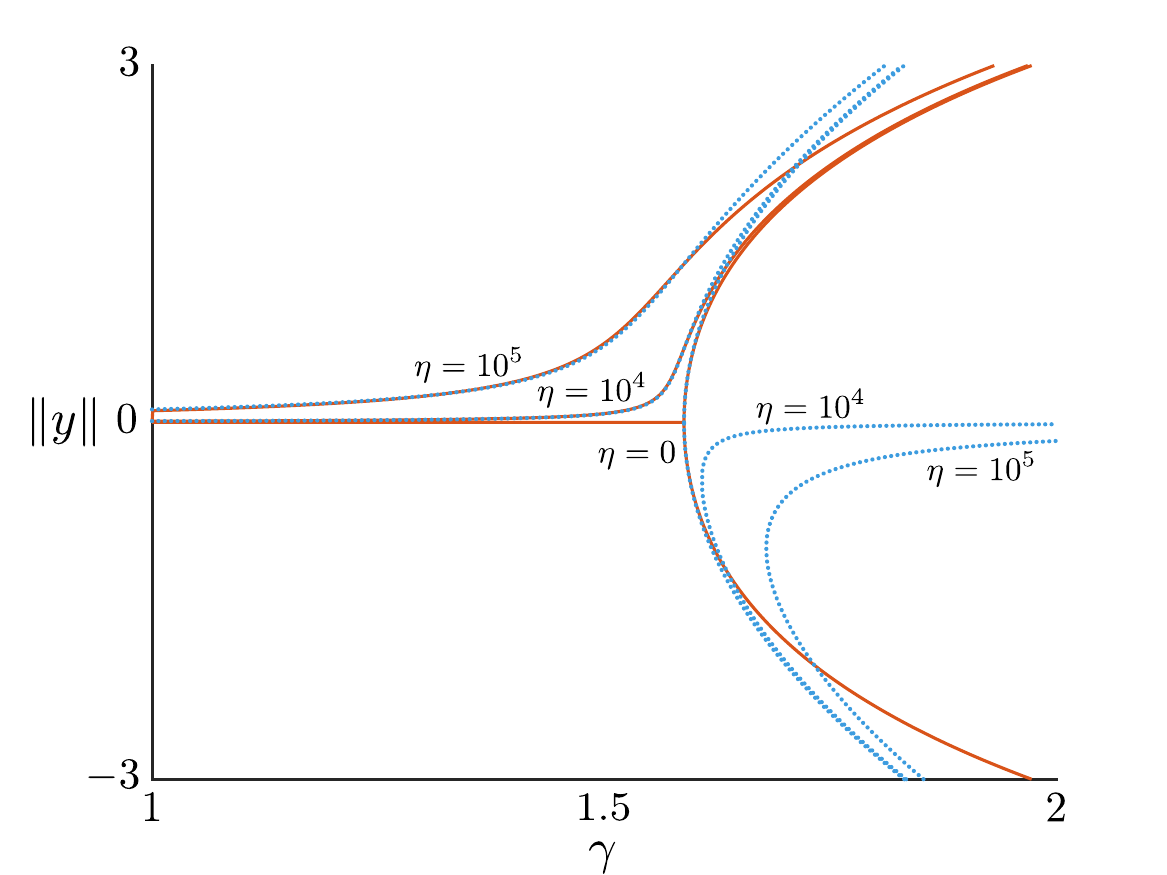}
\caption{\textbf{Bifurcation diagram for additive foundation heterogeneity \eqref{eq:forceeqnsimperfectfoundation}}. The foundation stiffness and rod length have been set to $\widehat{k} = 0.04$ and $\widehat{L}_0 = 20$, respectively, while the foundation heterogeneity $\xi(S_0)$ has been prescribed to $\xi(S_0) = \widehat{y}(S_0)$, as defined by Equation \eqref{eq:yhatsolution}. The amplitudes $\pm\|y\| = \pm\max_{S_0}|y(S_0)|$ against $\gamma$ from weakly nonlinear analysis (blue, dotted lines) and numerical continuation (red, solid lines) have been plotted for different values of $\eta$. For numerical calculations, the small parameter $\delta$ is set to $\delta = 0.01$ for all cases. Increasing $\eta$ further biases the rod to $\gamma = 1$ to the upper non-trivial amplitude branch. }
\label{fig:foundationheterogeneity}
\end{figure}

In order to produce a non-zero added term $K_4$, independent of $C_1$, we must consider {\it additive} heterogeneity. One possible type of additive heterogeneity, commonly considered in imperfection analyses, is in the shape of the foundation; that is, we consider the foundation to have spatially-varying imperfections present. That is, we modify the force balance equations \eqref{eq:forceeqnsnondimsimplified} to
\begin{align}
\frac{\partial F}{\partial S_0} = \widehat{k}(x - S_0), \qquad \frac{\partial G}{\partial S_0} = \widehat{k}(y - \widehat{\epsilon}\xi(S_0)),\label{eq:forceeqnsimperfectfoundation}
\end{align}
where $\xi(S_0)$ is the shape of the imperfection and $\widehat{\epsilon}$ captures the magnitude. This form of heterogeneity first affects the weakly nonlinear analysis when $\widehat{\epsilon} = O(\delta^3)$. Setting $\widehat{\epsilon} = \delta^3\eta$, where $\eta$ acts as a control parameter away from homogeneity, the solvability condition \eqref{eq:y3solvability} is shifted by a factor independent of $C_1$, and the bifurcation now satisfies
\begin{align}
K_1C_1^3 + K_2C_1\gamma^{(1)} + K_4\eta = 0.\label{eq:bucklingamplitudeeqnimperfectfoundation}
\end{align}
The constant $K_4$ is defined by
\begin{align}
K_4 = \widehat{k}\gamma_0^*\int^{\widehat{L}_0}_0\xi\widehat{y}dS_0.
\end{align}
(The constants $K_1$ and $K_2$ are the same as in the homogeneous case \eqref{eq:bucklingamplitudeeqn}.) Note the loss of both the trivial amplitude branch and the symmetric nature of the non-trivial amplitude branches. Therefore, with this underlying imperfection, the model undergoes an asymmetric (or imperfect) pitchfork bifurcation, with the branch selected determined by the sign of $K_4$. As $\eta$ is increased, so is the deviation from the homogeneous amplitude equation \eqref{eq:bucklingamplitudeeqn}, and hence the splitting of the initially-symmetric non-trivial branches is amplified. 
Note also that $K_4$ involves a simple inner product with $\xi$ and the buckling mode $\widehat y$. The heterogeneity thus has maximum effect when the imperfection to the foundation is of the same shape as the buckling mode, i.e. when $\xi\propto\widehat y$.
Figure \ref{fig:foundationheterogeneity} displays the bifurcation diagram for $\pm\|y\|$ against $\gamma$ for various values of $\eta$, both from the weakly nonlinear analysis (dashed curves) and numerical solution of the full system (solid curves). We have taken the heterogeneity $\xi(S_0) = \widehat{y}(S_0)$. As expected, increasing $\eta$ further splits the branches and increases the predisposition to the upper branch\footnote{Setting $\xi(S_0) = -\widehat{y}(S_0)$ biases the buckled rod to the lower amplitude branch, as seen by setting $D_1=-1$.  (Note that numerical continuation, increasing $\gamma$ from the flat rod state, cannot be used to detect the split branches.)}.  Due to the scaling $\widehat{\epsilon}=\delta^3$, large values of $\eta$ are needed to observe a noticeable difference in the bifurcation. In Figure \ref{fig:foundationheterogeneity} we have taken $\eta = O(10^5)$, which grossly violates the asymptotic assumption that $\eta=O(1)$; since in this plot $\delta=10^{-2}$, even with $\eta=10^5$, $\widehat{\epsilon} =10^{-1}$, i.e. the perturbation is still small, and we find that the weakly nonlinear analysis matches the full model reasonably well. Also evident is that as heterogeneity is increased, a larger deflection is observed for a given $\gamma$. This is consistent with typical results that imperfection deforms load-deflection curves so that higher deflections occur under smaller loads \cite{Hutchinson1970}.

As a final point of interest, we wish to measure which form of heterogeneity has the greatest impact on the bifurcation. As a `metric' for comparison, following the typical engineering analysis of load-deflection, here we consider the compressive force as a function of growth. To examine this, we define the \emph{net axial stress}
\begin{align}
\overline{n_3} = \frac{1}{\widehat{L}_0}\left|\int^{\widehat{L}_0}_0n_3 dS_0\right| = \frac{1}{\widehat{L}_0}\left|\int^{\widehat{L}_0}_0F\cos\theta+G\sin\theta dS_0\right|.\label{eq:averageaxialforce}
\end{align}
\begin{figure}[t]
\centering
\hspace*{0.25cm}
\captionsetup{width=\textwidth}
\includegraphics[width=0.875\textwidth]{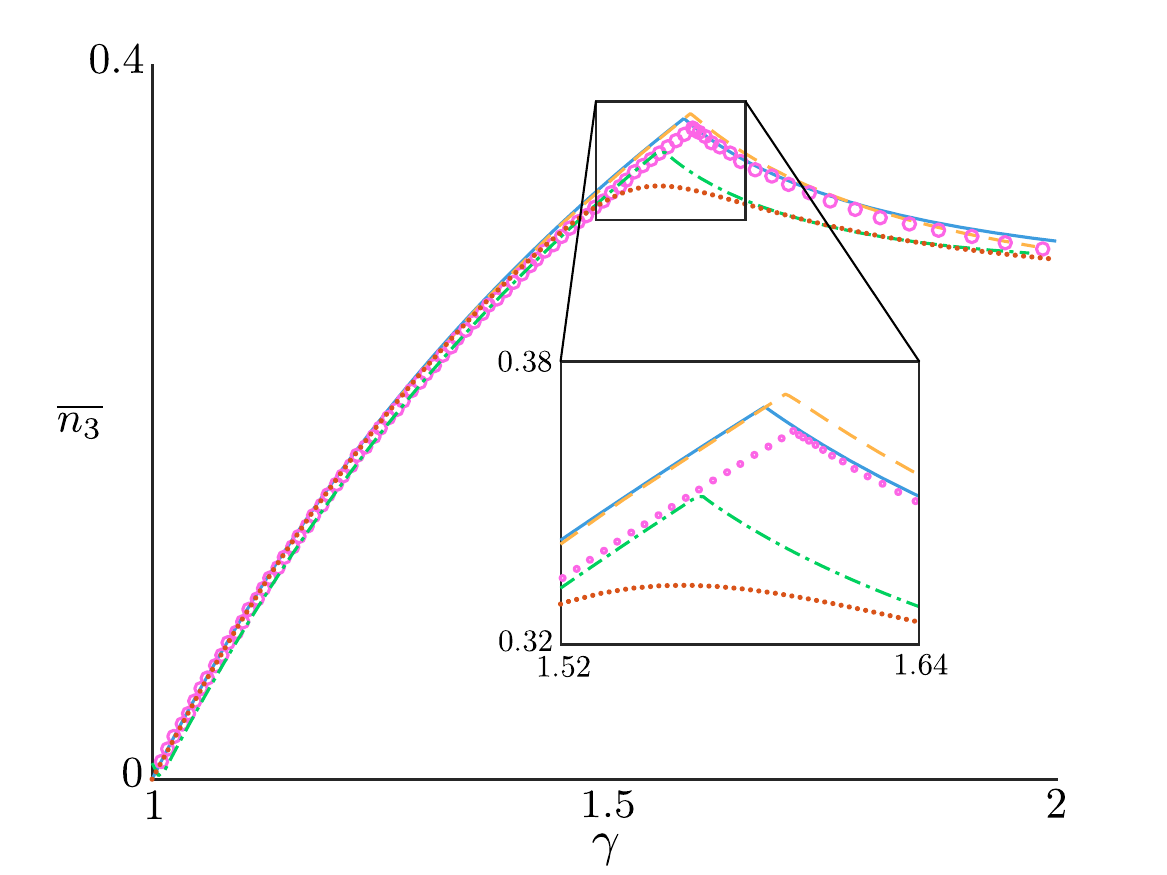}
\caption{\textbf{Effect of the considered heterogeneities on compressive stress}. The net axial load $\overline{n_3}$ is plotted against $\gamma$. As before, the foundation stiffness and rod length have been set to $\widehat{k} = 0.04$ and $\widehat{L}_0 = 20$, respectively, while the foundation heterogeneity $\xi(S_0)$ has been prescribed to $\xi(S_0) = \widehat{y}(S_0)$ (see Eq. \eqref{eq:yhatsolution}). The homogeneous case (blue, solid line) is compared against the foundation stiffness heterogeneity (orange, dashed lines); rod stiffness heterogeneity (pink, open circles); growth heterogeneity (green, dash-dotted lines); and foundation imperfection \eqref{eq:forceeqnsimperfectfoundation} (brown dots). For numerical calculations, the small parameter $\delta$ is set to $\delta = 0.01$ and $\eta$ is set such that $\widehat{\epsilon} = 0.1$ for all cases. The additive foundation heterogeneity is quickest to relieve the axial stress induced by rod growth. }
\label{fig:heterogeneitycomparison}
\end{figure}

In Figure \ref{fig:heterogeneitycomparison}, we compare the net axial stress with increasing growth for each of the four heterogeneities considered. In each case we have imposed $\xi=\hat y$, and due to the different nature of the perturbation schemes, we have chosen the scale factors such that the total perturbation from the uniform state is equivalent across the four cases. The perfect buckling case appears as the solid blue line, with the sharp cusp appearing at $\gamma^*$ and signifying that buckling occurs at a critical compressive stress, which is relieved partially through the buckling. Each of the heterogeneities produces a similar curve, though we see that the additive heterogeneity in foundation shape just considered has the most significant effect, followed by growth heterogeneity. Both foundation and stiffness heterogeneity follow the perfect case very closely; zooming in on the cusp region (see inset) shows that these forms lead to a delayed bifurcation, and, in the case of foundation stiffness, the bifurcation occurs at slightly larger stress before subsequently compensating and dipping below the perfect case.

\section{Discussion}
\label{sec:discussion}
We have investigated the buckling and post-buckling behaviour of a planar morphoelastic rod attached to an elastic foundation. 
We extended the original linear stability analysis by Moulton et al. \cite{Moulton2013} by conducting a weakly nonlinear analysis, complemented with numerical solutions of the full, nonlinear model. We first considered a homogeneous setting, and then explored the effect of heterogeneity in material parameters.

In the homogeneous case, we obtained a classic pitchfork bifurcation, with buckling occurring at a critical growth. The nature of the bifurcation (its location and type---supercritical or subcritical) could be characterised via two dimensionless parameters, one ($\widehat{L}_0$) relating to length of the finite rod, and another ($\widehat{k}$) comparing the relative stiffness of foundation and rod. Increasing length was found to destabilise the rod, causing bifurcation at a smaller value of growth and with increased mode number. Increasing the foundation stiffness, on the other hand, stabilises the rod, increasing the critical growth and the mode number.  \dem{The influence of foundation stiffness on the buckling mode and the onset of instability shows how variations in system parameters, even in a heterogeneous setting, can have a dramatic impact on the resulting morphology. Such results may have strong relevance in biological systems, where the precise form of the structure, e.g. number of folds, is crucial to functionality. A telling example is in the gyrification of the brain, where deviations in the developmental timing or degree of cortical folding have a severe neurological impact \cite{Goriely:2015cg}.} 

The general trends we have found in the homogeneous case are consistent with previous analyses of a similar nature, e.g. \cite{OKeeffe:2013et}. The type of bifurcation, however, was non-standard: the boundary between supercritical and subcritical bifurcations exhibited an unexpected complexity. In a biological context, where monotonically increasing growth is a natural driver of the formation and subsequent evolution of spatial patterns, this transition has critical importance, signifying where a smooth shape evolution (supercritical) can be expected \dem{as opposed to a discontinuous jump from a flat state (subcritical). While further work is needed to establish how prevalent the latter may be, the appearance of structural patterns that arise rapidly, such as the sharp spines in certain mollusc seashells \cite{Chirat2013} that appear directly adjacent to a flat portion of shell, may point to subcritical regimes.}
Here the effect of a finite domain is also apparent, as the complexity of the transition becomes less pronounced as $\widehat{L}_0$ increases.

Multiplicative heterogeneity with respect to three different material properties was then considered: the foundation stiffness, the rod stiffness, and growth. A modified weakly nonlinear analysis showed that in each case the heterogeneity served to translate the bifurcation point, but did not alter its nature. Explicit relations for the shift in bifurcation allowed us to determine how the form of the heterogeneity influences the direction and degree of the translation. For example, the simplest relation appeared with heterogeneity in the foundation stiffness, in which case the greatest effect occurs when the heterogeneity is aligned with the square of the buckling mode. This reflects the intuitive notion that weakening the foundation attachment in regions where the uniform rod deforms maximally has the strongest impact.

To complement the weakly nonlinear analysis, the full nonlinear system was solved with numerical continuation; this enabled us to investigate the post-buckling behaviour for more pronounced heterogeneity and at large growth values. 
A common feature was an induced `asymmetry' of the buckled shape. With heterogeneous foundation stiffness, softer (stiffer) parts of the foundation give rise to increased (decreased) rod amplitudes, as might be expected. With heterogeneity in rod stiffness, the situation is less straightforward. Softer parts of the rod are more easily curved, and thus it might be reasonable to expect such regions to correspond to higher amplitude; however, compression is also less costly in the soft regions. In all cases we have examined, the rod flattens through compression in the soft regions, a deformation that increases stretching energy, but is compensated by a decrease in both bending and foundation energies. Here, the assumption of extensibility is crucial, as compression is not permitted in an inextensible model. Indeed, a direct comparison of an inextensible and extensible model post-buckling revealed significant morphological differences, highlighting the importance of a critical assessment of when the inextensible assumption is warranted. In the case of non-homogeneous growth, we showed that even with zero net growth, heterogeneity, interpreted as a redistribution of rod material from spatial regions with decreased $\gamma$ to those with increased $\gamma$, can significantly impact the post-buckling behaviour. The general trend is not surprising: the rod flattens in regions where material is lost. What is perhaps surprising is that the distribution of material seems to play as important a role in the shape evolution as the total amount of material added through growth. 

\begin{figure}[t!]
\centering
\captionsetup{width=\textwidth}
\includegraphics[width=1.025\textwidth]{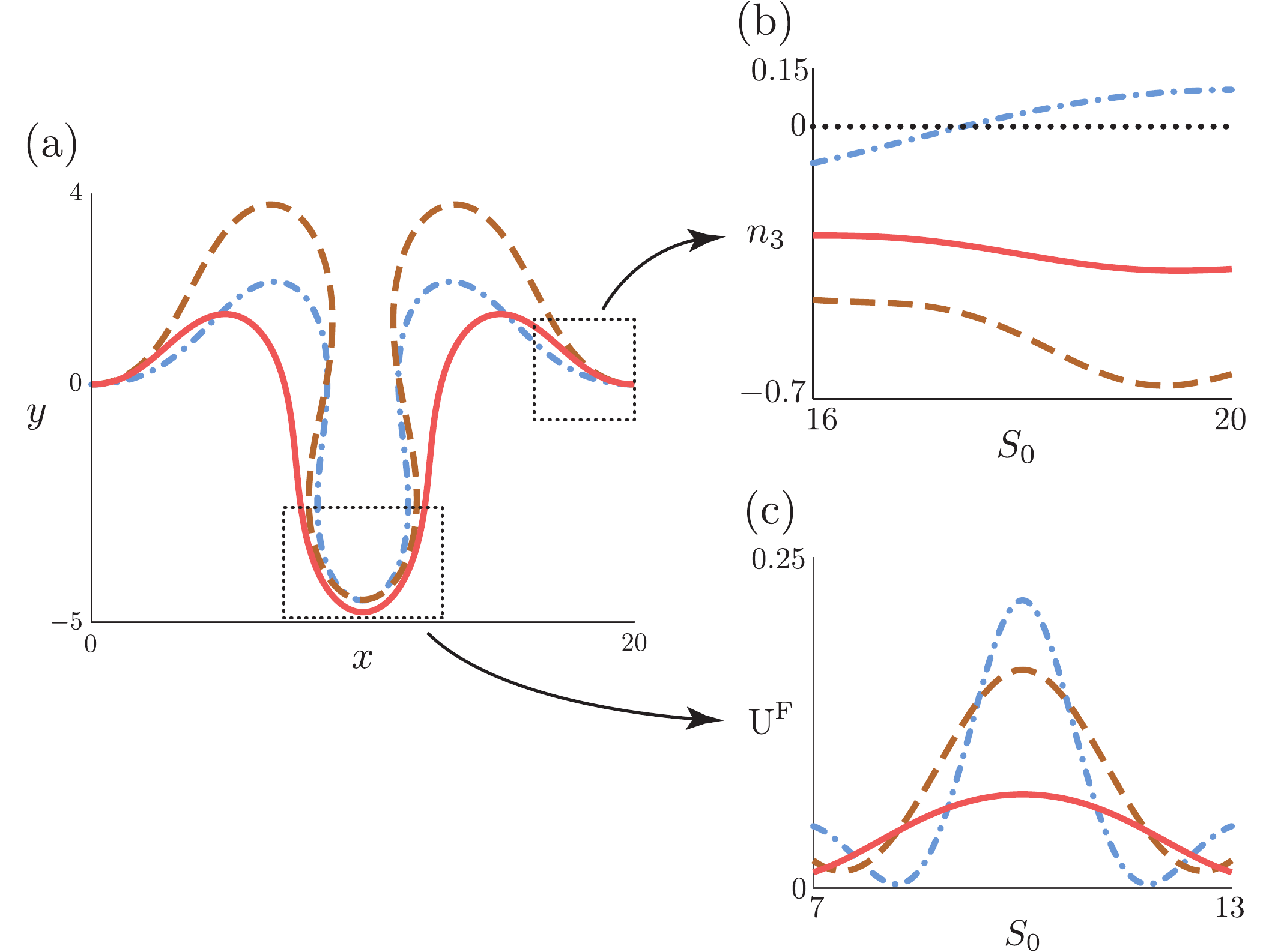}
\caption{\textbf{Inverse problem example.} (a) Different forms of heterogeneity in rod stiffness (brown, dashed), and growth (blue, dot-dashed) are chosen to approximately match the morphology produced by foundation stiffness heterogeneity (red, solid). (b) Despite the similar shape and amplitude in the edge  region, growth heterogeneity gives rise to tension ($n_3 > 0$). (c) Foundation stiffness heterogeneity decreases the foundation energy density $\mathrm{U}^\mathrm{F}$ in the central region where the foundation has been weakened.}
\label{fig:heterogeneitysignatures}
\end{figure}

\dem{Having examined the effect of heterogeneity on post-buckled shape, two natural and related questions follow from this: (i) can one tailor the heteregeneities to achieve a desired shape? And (ii) given a particular shape, can one infer the form and type of any material heterogeneity present? These questions, with significant relevance both from morphogenetic and tissue engineering perspectives, are related to the mathematical inverse problem. Such a problem is inherently complex, as the shapes considered are (partial) solutions of a high-order nonlinear boundary value problem, only achieved in the forward direction through numerical path continuation. In order to develop some intuition, here we provide a simple but illustrative example: we take a candidate shape with embedded heterogeneity---foundation heterogeneity following Figure \ref{fig:heterogeneitysummarised}(b)---and we try to match that shape, in a trial-and-error manner, by varying the heterogeneity in either the rod stiffness or the growth (as well as the net growth), guided by the results of Section \ref{sec:parameterheterogeneity}; we then consider characteristics other than the shape itself and seek distinctive differences, i.e. {\it signatures} of the heterogeneities (more details provided in Appendix \ref{sec:inverseproblem}).}   

\dem{The result of this exercise is summarised in Figure \ref{fig:heterogeneitysignatures}. In Figure \ref{fig:heterogeneitysignatures}(a) we plot the `matched' shapes; clearly the match is imperfect, highlighting already the non-trivial nature of tailoring the heterogeneity to achieve a specific shape. Figure \ref{fig:heterogeneitysignatures}(b) shows the axial stress $n_3$ in the outer region. We observe that growth heterogeneity produces disparate regions of tension, $n_3 > 0$ (where growth is reduced), and compression, $n_3 < 0$ (where growth is increased). In contrast, the rod remains in compression for foundation and rod stiffness. In distinguishing foundation stiffness heterogeneity, the foundation energy density $\mathrm{U}^\mathrm{F}$, defined explicitly in Appendix \ref{sec:energy}, provided the clearest indicator. Figure \ref{fig:heterogeneitysignatures}(c) plots $\mathrm{U}^\mathrm{F}$ in the middle section of the rod, where the shapes are qualitatively most similar, and we find a significant decrease in the case of foundation heterogeneity.}

In a thought experiment where the morphology is given and the task is to determine the heterogeneity, these differences could in principle be detected by cutting experiments that release residual stress, as is done for instance in arteries \cite{chuong1986residual} and solid tumours \cite{stylianopoulos2012causes}. However, while this example suggests the possibility of distinguishing between forms of heterogeneity and using heterogeneity to tailor properties, it is clear that this is not a straightforward problem, and a more rigorous treatment would be needed to reach firm conclusions. Moreover, we observed no features that clearly distinguished the case of rod stiffness heterogeneity from the other two heterogeneities. In a 3D setting, more measurable quantities are available, for example, stress in transverse directions, which could potentially yield measurable differences in behaviour. On the other hand, the general complexity of the inverse problem will increase as the number of variables increases. In any case, modelling studies and computational and/or analytical results such as those provided by a weakly nonlinear analysis can provide important insights in a tissue engineering context, e.g. determining the right `ingredients' to generate desired tissue morphologies; as well as for building intuition for how different regions of a heterogeneous elastic tissue with evolving material properties will behave. For instance, this is of particular relevance in brain injury, where morphological heterogeneities have crucial influence in understanding the deformation response to injury \cite{Goriely:2015cg}, and in intestinal tissue health, where deformation plays a significant role in facilitating wound healing \cite{seno2009efficient} and tumour expansion \cite{Preston2003}.

In the final section, we have examined a fourth type of heterogeneity, with a view to establishing why the impact of heterogeneity on the bifurcation itself was relatively minor in the previous scenarios. Here we made the key distinction between multiplicative and additive heterogeneity. A multiplicative heterogeneity appears in a term that multiplies dependent variables in the system; due to the nature of the perturbative expansion, such terms only serve to shift the pitchfork bifurcation. An additive heterogeneity, for which a perturbation is applied to a term that does not multiply dependent variables, can have a significant effect, creating an imperfect bifurcation (broken pitchfork) and creating a larger deviation from the perfect, homogeneous, case. Here we considered an imperfectly-straight foundation, and showed that the effect is maximal when the form of the imperfection matches that of the buckling mode.

In this paper, we have assumed each model parameter to be independent from the others and, for the sake of clarity, varied each parameter in isolation. A natural extension would be to consider the combined effect of several simultaneously-varying parameters via inter-parameter coupling.  For example, one could consider a rod with non-uniform stiffness and a growth evolution law that depends on axial stress. This would naturally induce heterogeneity in multiple parameters, and the resultant competing effects would likely produce a complex solution space.  Another natural extension is to consider nonlinear constitutive effects. There would certainly be benefit to considering a nonlinear constitutive relation between axial stress and the elastic stretch $\alpha$, in particular because many of our simulations featured significant compression potentially beyond the threshold for quantitative validity of the Hooke's law considered here. Another useful extension is to incorporate nonlinearity in the response of the foundation, a phenomenon that has been studied in great detail in systems without growth \cite{Hutchinson2013}, but whose role in the context of growth remains unclear. 

Finally, we note that many of our modelling choices were motivated by observations on the intestinal crypt (and other, physiologically similar structures). Thus while the model represents an idealised version of a crypt, there are several extensions that would render it biologically realistic. For instance, our growth parameter contains no information about the timescale of growth, which is a fundamental aspect of many biological systems, particularly the crypt. Therefore, one could introduce time-dependent growth or time-dependent mechanical relaxation (for example in the foundation), allowing remodelling to occur over time. Alternatively, the proliferative structure within a crypt suggests the spatial form of the growth stretch should be bimodal \cite{Alberts2017,Wright1984}. In the context of mechanosensitive growth, Miyoshi et al.\ \cite{Miyoshi2012} showed that a specific subset of stromal cells is activated during wound healing to increase stem cell proliferation in the crypt, as one example. The crypt also provides a natural setting to investigate possible feedback mechanisms between growth and the underlying foundation; this work is currently underway.

\section*{Acknowledgements}
This work was supported by Cancer Research UK (CRUK) grant number C5255/A23225, through a Cancer Research UK Oxford Centre Prize DPhil Studentship. PKM would like to thank the Mathematical Biosciences Institute (MBI) at Ohio State University, for partially supporting this research. MBI receives its funding through the National Science Foundation grant DMS1440386. The authors thanks A. Goriely for useful discussions.

\bibliographystyle{spmpsci}   
\bibliography{PlanarMorphorodsWNA.bib}

\appendix
\section{Determining the buckling amplitude}
\label{sec:bucklingamplitudec1}
In this section, we introduce the functions that are needed to calculate the buckling amplitude $C_1$ for different values of $\widehat{k}$ and $\widehat{L}_0$ (and hence $\gamma^*$). After unfolding the bifurcation with the ansatz \eqref{eq:growthunfolding} and considering $\mathcal{O}(\delta^2)$ terms, we obtain the system
\begin{align}
&x^{(2)'}-\gamma^*F^{(2)} = \frac{\gamma^{(1)}}{\gamma^*} - \frac{\gamma^*}{2}\left(y^{(1)'}\right)^2 - y^{(1)'}y^{(1)'''},\\
&y^{(2)'} - \theta^{(2)} = 0,\\
&F^{(2)'} - \widehat{k}x^{(2)} = 0,\\
&G^{(2)'} - \widehat{k}y^{(2)} = 0,\\
&\theta^{(2)'} - \gamma^*m^{(2)} = 0,\\
&m^{(2)'} + \frac{(\gamma^*-1)}{\gamma^*}\theta^{(2)} + G^{(2)} = 0.\label{eq:moment2}
\end{align}
We note that we have used Equations \eqref{eq:geomeqnslinearised}--\eqref{eq:momenteqnslinearised} to express the inhomogeneities in terms of $y^{(1)}$ only, and have simplified the system further by substituting $x^{(1)} = F^{(1)} = 0$. As in Section \ref{sec:linearstabilityanalysis}, the system decouples into two ordinary differential equations for $x^{(2)}$ and $y^{(2)}$ given by \eqref{eq:xdirectioneqngeneraln}--\eqref{eq:ydirectioneqngeneraln}, with
\begin{align}
Lx^{(2)} = H_{x^{(2)}} &= -\gamma^*y^{(1)'}y^{(1)''}-\left(y^{(1)'}y^{(1)'''}\right)',\label{eq:xdirectioneqn2}\\
My^{(2)} = H_{y^{(2)}}  &= 0. \label{eq:ydirectioneqn2}
\end{align}
We recall the linear operators $L$ and $M$ are defined in Equations \eqref{eq:xdirectioneqngeneraln} and \eqref{eq:ydirectioneqngeneraln} respectively by $Lx = x'' - \widehat{k}\gamma^*x$ and $My = y'''' + (\gamma^*-1)y'' + \widehat{k}\gamma^*y$. As $H_{y^{(2)}} = 0$, \eqref{eq:ydirectioneqngeneraln} is homogeneous for $n = 2$ and provides no information about the buckling amplitude $C_1$. However, considering Equations \eqref{eq:xdirectioneqngeneraln} and \eqref{eq:xdirectioneqn2} in tandem with the solution \eqref{eq:y1solution} and the boundary conditions $x^{(2)}(0) = x^{(2)}(\widehat{L}_0) = 0$ allows us to define $\widehat{x} :=C_1^{-2}x^{(2)}$.

At next order, we have 
\begin{align}
&x^{(3)'}-\gamma^*F^{(3)} = -\gamma^*y^{(1)'}y^{(2)'} - y^{(1)'}y^{(2)'''} - y^{(1)'''}y^{(2)'},\label{eq:geomeqns3}\\
&y^{(3)'} - \theta^{(3)} = \frac{1}{3}\left(y^{(1)'}\right)^3 + x^{(2)'}y^{(1)'},\\
&F^{(3)'} - \widehat{k}x^{(3)} = 0,\\
&G^{(3)'} - \widehat{k}y^{(3)} = 0,\\
&\theta^{(3)'} - \gamma^*m^{(3)} = \frac{\gamma^{(1)}}{\gamma^*}y^{(1)''},\\
&m^{(3)'} + \frac{(\gamma^*-1)}{\gamma^*}\theta^{(3)} + G^{(3)} = \frac{(\gamma^*+2)}{6\gamma^*}\left(y^{(1)'}\right)^3 + \frac{1}{\gamma^*}\left(y^{(1)'}\right)^2y^{(1)'''} + \frac{1}{\gamma^*}x^{(2)'}y^{(1)'}\nonumber
\\& \qquad\qquad\qquad\qquad\qquad\qquad + \frac{1}{\gamma^*}x^{(2)'}y^{(1)'''} - \frac{\gamma^{(1)}}{\gamma^{*^2}}y^{(1)'}.\label{eq:momenteqns3}
\end{align}
Equations \eqref{eq:geomeqns3}--\eqref{eq:momenteqns3} decouple into the two ordinary differential equations, in which the forcing terms $H_{x^{(n)}}$ and $H_{y^{(n)}}$ are given by:
\begin{align}
Lx^{(3)} = H_{x^{(3)}} &= -\left(\gamma^*y^{(1)'}y^{(2)'} + y^{(1)'''}y^{(2)'} + y^{(1)'}y^{(2)'''}\right)',\label{eq:x3inhomogeneities}\\
My^{(3)} = H_{y^{(3)}} &= 2\left(y^{(1)''}\right)^3 + \frac{3\gamma^*}{2}\left(y^{(1)'}\right)^2y^{(1)''} + 2\left(y^{(1)'}\right)^2y^{(1)''''} + 8y^{(1)'}y^{(1)''}y^{(1)'''}\nonumber
\\& + \gamma^*\left(x^{(2)'}y^{(1)'}\right)' + y^{(1)'}x^{(2)''''} + 2x^{(2)'}y^{(1)''''} + 4x^{(2)''}y^{(1)'''}\nonumber
\\&+ 3x^{(2)'''}y^{(1)''} - \frac{\gamma^{(1)}}{\gamma^*}\left(y^{(1)''} - y^{(1)''''}\right).\label{eq:y3inhomogeneities}
\end{align}
Using the Fredholm Alternative Theorem and considering \eqref{eq:y3inhomogeneities} in powers of $C_1$, the constants $K_1$ and $K_2$ in Equation \eqref{eq:bucklingamplitudeeqn} are obtained by evaluating the following integrals
\begin{align}
K_1 &= \int^{\widehat{L}_0}_0\Bigg[2\left(\widehat{y}^{''}\right)^3 + \frac{3\gamma^*}{2}\left(\widehat{y}^{'}\right)^2\widehat{y}^{''} + 2\left(\widehat{y}^{'^2}\widehat{y}^{'''}\right)'+ 4\widehat{y}^{'}\widehat{y}^{''}\widehat{y}^{'''}\nonumber\\
&\qquad\qquad + \gamma^*\left(\widehat{x}^{'}\widehat{y}^{'} \right)' + 2\left(\widehat{x}^{'}\widehat{y}^{''}\right)'' + \left(\widehat{x}^{'''}\widehat{y}^{'}\right)' \Bigg]\widehat{y}dS_0,\label{eq:K1C1}\\
K_2 &= -\frac{1}{\gamma^*}\int^{\widehat{L}_0}_0\left(\widehat{y}^{''} - \widehat{y}^{''''}\right)\widehat{y}dS_0,\label{eq:K2C1}
\end{align}
where $\widehat{x} := C_1^{-2}x^{(2)}$ and $\widehat{y} := C_1^{-1}y^{(1)}$. Applying integration by parts to \eqref{eq:K2C1} yields
\begin{align}
K_2 = \frac{1}{\gamma^*}\int^{\widehat{L}_0}_0\left(\widehat{y}^{'}\right)^2 + \left(\widehat{y}^{''}\right)^2dS_0.\label{eq:K2C1simplified}
\end{align}
Therefore $K_2 > 0$ for all parameter values, provided that $\widehat{y}$ is non-trivial. Hence, the sign of constant $K_1$ determines the nature of the pitchfork. 
\section{Pitchfork bifurcation on an infinite domain}
\label{sec:pitchforkinfinitedomain}
We show that in the case of an infinitely-long rod, the value of $\widehat{k}$ for which the system transitions from a supercritical pitchfork bifurcation to a subcritical pitchfork bifurcation can be calculated exactly.  Moreover, we show that the transition occurs only once.

Recall that the linear stability analysis yields the ordinary differential equations equations (Equations \eqref{eq:xdirectioneqnlinearised}--\eqref{eq:ydirectioneqnlinearised})
\begin{align}
Lx^{(1)} = x^{(1)''} - k\gamma x^{(1)} = 0, \qquad My^{(1)} = y^{(1)''''} + (\gamma - 1)y^{(1)''} + k\gamma y^{(1)} = 0,
\end{align}
Seeking oscillatory modes $y^{(1)} \sim e^{i\omega S_0}$ yields the oscillation frequencies from Equation \eqref{eq:omega1and2}, $\omega^2_{\pm} = \frac{\gamma - 1}{2} \pm \left(\frac{(\gamma - 1)^2}{4} - \widehat{k}\gamma\right)^{\frac{1}{2}}$. For oscillations to persist over the whole domain, we require that $(\gamma - 1)^2 \ge 4\widehat{k}\gamma$. Hence the bifurcation occurs when $(\gamma - 1)^2 = 4\widehat{k}\gamma$, which gives rise to $\gamma^*_\mathrm{inf}$, defined by Equation \eqref{eq:gammazero}. Enforcing boundedness and that the solution is real-valued leads to the solution
\begin{align}
y^{(1)} = C_1\cos(\omega S_0), \label{eq:y1infdomain}
\end{align}
where
\begin{align}
\omega = \left(\frac{\gamma^*_{\mathrm{inf}} - 1}{2}\right)^{\frac{1}{2}} = \left(\widehat{k} + \left(\widehat{k} + \widehat{k}^2\right)^{\frac{1}{2}}\right)^{\frac{1}{2}}.\label{eq:omegainfdomain}
\end{align}
As before, we unfold about the bifurcation point $\gamma^*_\mathrm{inf}$ by letting $\gamma = \gamma^*_\mathrm{inf} + \delta^2\gamma^{(1)}$, where $\delta \ll 1$ is our perturbation expansion parameter, and $\gamma^{(1)}$ is the control parameter away from bifurcation.

 At $O(\delta^2)$, we substitute \eqref{eq:y1infdomain} into \eqref{eq:xdirectioneqn2} and, after solving the equation and imposing boundedness, obtain 
\begin{align}
x^{(2)} = C_1^2\frac{\omega^3(\gamma^*_\mathrm{inf} - 2\omega^2)}{8\omega^2 + 2\widehat{k}\gamma^*_\mathrm{inf}}\sin(2\omega S_0). \label{eq:x2infdomain}
\end{align}
At $O(\delta^3)$, we obtain the amplitude equation for $C_1$ by substituting \eqref{eq:y1infdomain} and \eqref{eq:x2infdomain} into \eqref{eq:y3inhomogeneities} and impose that secular terms vanish. This leads to Equation \eqref{eq:bucklingamplitudeeqn}, $K_1C_1^3 + K_2C_1\gamma^{(1)} = 0$, where the constants $K_1$ and $K_2$ are given by
\begin{align}
K_1 &= \frac{\widehat{k}^2\left[7 + 100 \left(\widehat{k} + \widehat{k}^2\right)^{\frac{1}{2}} + 8\widehat{k}\left(16 + 7\widehat{k} + 7 \left(\widehat{k} + \widehat{k}^2\right)^{\frac{1}{2}}\right)\right] - 4\widehat{k}\left(8 + 9 \left(\widehat{k} + \widehat{k}^2\right)^{\frac{1}{2}}\right)}{8\left(16 + 7\widehat{k}\right)},\label{eq:K1infdomain}
\\K_2 &= \left(\widehat{k} + \widehat{k}^2\right)^{\frac{1}{2}}.\label{eq:K2infdomain}
\end{align}
We note that we have substituted Equation \eqref{eq:gammazero} to express $K_1$ and $K_2$ in terms of $\widehat{k}$ only. As $K_2>0\ \forall \ \widehat{k} > 0$, the sign of $K_1$ completely determines the type of pitchfork bifurcation the system undergoes, as confirmed in Appendix \ref{sec:bucklingamplitudec1}. Therefore, the value of $\widehat{k} > 0$ where the transition occurs is found by solving $K_1 = 0$, yielding
\begin{align}
\widehat{k}^* = \frac{21\sqrt{89} - 137}{160} \approx 0.38196. \label{eq:criticalkinfdomain}
\end{align}
It can be verified easily that $K_1 < 0$ when $\widehat{k} < \widehat{k}^*$ and $K_1 > 0$ when $\widehat{k} > \widehat{k}^*$, indicating a transition from a supercritical pitchfork bifurcation to a subcritical pitchfork bifurcation. Moreover, there is only one transition when the rod is of infinite length, in contrast to the multiple transitions that can occur when the rod length is finite.
\section{Including parameter heterogeneity}
\label{sec:heterogeneityfunctions}
Here, we list the additional inhomogeneities that appear in Equations \eqref{eq:geomexpandedn}--\eqref{eq:momentgeneraln} when we account for spatial heterogeneities. These inhomogeneities result in the subsequent change from the buckling amplitude equation for the homogeneous case \eqref{eq:bucklingamplitudeeqn} to Equation \eqref{eq:bucklingamplitudeeqnvariablefoundation}.

In the first case of foundation stiffness heterogeneity, the linearised equations at $O(\delta)$ are unchanged from the homogeneous case. At $O(\delta^2)$ and $O(\delta^3)$, we have the following system of equations:
\begin{align}
&Lx^{(2)} = H_{x^{(2)}},\label{eq:xdirectioneqn2variablefoundation}\\
&My^{(2)} = 0,\label{eq:ydirectioneqn2variablefoundation}\\
&Lx^{(3)} = H_{x^{(3)}},\label{eq:xdirectioneqn3variablefoundation}\\
&My^{(3)} = H_{y^{(3)}} -\eta \widehat{k}_0{\gamma_0^*}\xi y^{(1)}.\label{eq:y3inhomogeneitiesvariablefoundation}
\end{align} 
The linear operators $L$ and $M$ have been defined in Equations \eqref{eq:xdirectioneqngeneraln}--\eqref{eq:ydirectioneqngeneraln} and Appendix \ref{sec:bucklingamplitudec1}, while the inhomogeneities $H_{x^{(2)}},\ H_{x^{(3)}},$ and $H_{y^{(3)}}$ are given by Equations \eqref{eq:xdirectioneqn2}, \eqref{eq:x3inhomogeneities} and \eqref{eq:y3inhomogeneities}. Thus the amplitude equation now satisfies \eqref{eq:bucklingamplitudeeqnvariablefoundation}, $K_3$ can be deduced by evaluating the solvability condition, \eqref{eq:y3solvability}.

When rod stiffness heterogeneity is considered, the equations at $O(\delta^2)$ are
\begin{align}
&Lx^{(2)} = H_{x^{(2)}} + \eta(\gamma^*_0 - 1)\xi',\label{eq:xdirectioneqn2variablerodstiffness}\\
&My^{(2)} = 0.\label{eq:ydirectioneqn2variablerodstiffness}
\end{align}
As Equation \eqref{eq:xdirectioneqn2variablerodstiffness} is a linear ordinary differential equation, the superposition of particular solutions and the boundary conditions $x^{(2)}(0) = x^{(2)}(\widehat{L}_0) = 0$ imply that we can define $x^{(2)} := C_1^2\widehat{x} + \eta\widehat{x}_\xi$, where $L\widehat{x}_\xi=(\gamma^*_0 - 1)\xi'$. At the next order, we have
\begin{align}
Lx^{(3)} &= H_{x^{(3)}}\label{eq:x3inhomogeneitiesvariableyoungsmodulus},\\
My^{(3)} &= H_{y^{(3)}} + \eta\bigg[\xi\left((1-\gamma^*_0)y^{(1)''} - y^{(1)''''}\right)\nonumber\\
&\quad+ \xi'\left((1-\gamma^*_0)y^{(1)'} - y^{(1)'''}\right) - \xi'y^{(1)'''} - \xi''y^{(1)''}\bigg].\label{eq:y3inhomogeneitiesvariableyoungsmodulus}
\end{align} 
Therefore, $K_3$ is obtained by evaluating \eqref{eq:y3solvability}, yielding the amplitude equation \eqref{eq:bucklingamplitudeeqnvariablefoundation}. 

For heterogeneous growth modelled by \eqref{eq:arbitraryheterogeneity}, considering $O(\delta^2)$ terms now yields
\begin{align}
Lx^{(2)} &=  H_{x^{(2)}} + \eta\xi',\label{eq:xdirectioneqn2variablegrowth}\\
My^{(2)} &=  0,\label{eq:xdirectioneqn2variablegrowth}
\end{align}
Therefore, by the linearity of \eqref{eq:xdirectioneqn2variablegrowth}, we can again write $x^{(2)} := C_1^2\widehat{x} + \eta\widehat{x}_\xi$ where, $x_\xi$ satisfies $L\widehat{x}_\xi=\xi'$. Furthermore, \begin{align}
Lx^{(3)} &= H_{x^{(3)}}\label{eq:x3inhomogeneitiesvariablegrowth},\\
M{y^{(3)}} &= H_{y^{(3)}} - \eta\left[\xi\left(y^{(1)''} + y^{(1)''''}\right) + \xi'\left(y^{(1)'} + y^{(1)'''}\right)\right].\label{eq:y3inhomogeneitiesvariablegrowth}
\end{align} 
Hence, evaluating \eqref{eq:y3solvability} for growth heterogeneity yields the amplitude equation \eqref{eq:bucklingamplitudeeqnvariablefoundation}, with $K_3$ obtained by the Fredholm Alternative Theorem. 

\section{Stability analysis of buckled solutions}
\label{sec:dynamicstability}
Despite the insights provided from the linear stability analysis and weakly nonlinear analysis on the growth-induced evolution of the rod, we have no information about the dynamic stability of the non-trivial solutions obtained, as we have only considered the static form of the model. In order to investigate the stability of the buckled solutions, we must consider the time-dependent behaviour of the system. We assume that there is no dynamic rotation in the system, and therefore only introduce time-dependence to the force balance. We scale time by the standard Kirchoff time scaling $T = (\rho I/EA)^{1/2}\widehat{T}$ \cite{Coleman1993,Goriely1997a,Goriely1997}. In the initial configuration, the planar force balance equations \eqref{eq:forceeqnsnondimsimplified} are
\begin{align}
&F' = \widehat{k}(x-S_0) + \gamma\ddot{x},\label{eq:xforceeqntimedependent}\\
&G' = \widehat{k}y + \gamma\ddot{y},\label{eq:yforceeqntimedependent}
\end{align}
where $' = \partial/\partial S_0$ $\dot{\ } = \partial/\partial\widehat{T}$. Suppose that a known equilibrium solution to the full nonlinear system \eqref{eq:geomeqnsnondimsimplified}--\eqref{eq:nondimBCs} is given by the solution vector 
\begin{align*}
\mathbf{x}_\mathrm{eq} = \left(x_{\mathrm{eq}}(S_0), y_\mathrm{eq}(S_0), \theta_\mathrm{eq}(S_0),F_\mathrm{eq}(S_0),G_\mathrm{eq}(S_0),m_\mathrm{eq}(S_0)\right)^T,
\end{align*}
 with fixed parameters $k,\ L$ and $\gamma$, and $T$ denotes the matrix transpose. We perform a linear time-dependent perturbation in the arbitrarily small parameter $\delta$ as such
\begin{align}
&x = x_{\mathrm{eq}}(S_0) + \delta x_{\mathrm{dyn}}(S_0)e^{i\sigma T},\nonumber\\
&y = y_{\mathrm{eq}}(S_0) + \delta y_{\mathrm{dyn}}(S_0)e^{i\sigma T},\nonumber\\
&\theta = \theta_{\mathrm{eq}}(S_0) + \delta \theta_{\mathrm{dyn}}(S_0)e^{i\sigma T},\nonumber\\
&F = F_{\mathrm{eq}}(S_0) + \delta F_{\mathrm{dyn}}(S_0)e^{i\sigma T},\nonumber\\
&G = G_{\mathrm{eq}}(S_0) + \delta G_{\mathrm{dyn}}(S_0)e^{i\sigma T},\nonumber\\
&m = m_{\mathrm{eq}}(S_0) + \delta m_{\mathrm{dyn}}(S_0)e^{i\sigma T}.\label{eq:timedependentperturbation}
\end{align}
Substituting this perturbation into \eqref{eq:geomeqnsnondimsimplified}, \eqref{eq:momenteqnsnondimsimplified}, \eqref{eq:xforceeqntimedependent}, and \eqref{eq:yforceeqntimedependent} yields the system, at $O(\delta)$:
\begin{align}
&x'_\mathrm{dyn} = \frac{\gamma}{2}\big(1 + \cos(2\theta_\mathrm{eq})\big)F_\mathrm{dyn} + \frac{\gamma}{2}\sin(2\theta_\mathrm{eq})G_\mathrm{dyn}\nonumber\\
&\quad\qquad - \gamma\big[\sin\theta_\mathrm{eq} + \sin(2\theta_\mathrm{eq})F_\mathrm{eq} - \cos(2\theta_\mathrm{eq})G_\mathrm{eq}\big]\theta_\mathrm{dyn},\label{eq:xtimepert}\\
&y'_\mathrm{dyn} = \frac{\gamma}{2}\sin(2\theta_\mathrm{eq})F_\mathrm{dyn} + \frac{\gamma}{2}\big(1-\cos(2\theta_\mathrm{eq})\big)G_\mathrm{dyn}\nonumber\\
&\quad\qquad - \gamma\big[\cos\theta_\mathrm{eq} + \cos(2\theta_\mathrm{eq})F_\mathrm{eq} - \sin(2\theta_\mathrm{eq})G_\mathrm{eq}\big]\theta_\mathrm{dyn},\label{eq:ytimepert}\\
&F'_\mathrm{dyn} = (\widehat{k} - \gamma\sigma^2)x_\mathrm{dyn},\label{eq:Ftimepert}\\
&G'_\mathrm{dyn} = (\widehat{k}-\gamma\sigma^2)y_\mathrm{dyn},\label{eq:Gtimepert}\\
&\theta'_\mathrm{dyn} = \gamma m_\mathrm{dyn},\label{eq:thetatimepert}\\
&m'_\mathrm{dyn} = \gamma\big[\sin\theta_\mathrm{eq}+F_\mathrm{eq}\sin(2\theta_\mathrm{eq})-G_\mathrm{eq}\cos(2\theta_\mathrm{eq})\big]F_\mathrm{dyn}\nonumber\\
&\quad\qquad - \gamma\big[\cos\theta_\mathrm{eq}+ F_\mathrm{eq}\cos(2\theta_\mathrm{eq})-G_\mathrm{eq}\sin(2\theta_\mathrm{eq})\big]G_\mathrm{dyn}\nonumber\\
&\quad\qquad + \gamma\Big[F_\mathrm{eq}\cos\theta_\mathrm{eq} + G_\mathrm{eq}\sin\theta_\mathrm{eq} + 2F_\mathrm{eq}G_\mathrm{eq}\sin(2\theta_\mathrm{eq})\nonumber\\
&\qquad\qquad\quad + (F^2_\mathrm{eq} - G^2_\mathrm{eq})\cos(2\theta_\mathrm{eq})\Big]\theta_\mathrm{dyn},\label{eq:momenttimepert}
\end{align}
We note that we have made use of the trigonometric double angle formulae to simplify the ordinary differential equations. The system \eqref{eq:xtimepert}--\eqref{eq:momenttimepert} can be written in the form
\begin{align}
\mathbf{x}_\mathrm{dyn}' = \mathbf{A}\mathbf{x}_\mathrm{dyn}, \label{eq:timepertsystem}
\end{align}
where the vector $\mathbf{x}_\mathrm{dyn} = (x_\mathrm{dyn},y_\mathrm{dyn},\theta_\mathrm{dyn},F_\mathrm{dyn},G_\mathrm{dyn}, m_\mathrm{dyn})^T$ and $\mathbf{A}$ is a matrix that is expressed solely in terms of $\mathbf{x}_\mathrm{eq}$. As $\mathbf{x}_\mathrm{eq}$ is known, for a given $\sigma^2$, $\mathbf{A}$ is constant for each $S_0 \in [0, \widehat{L}_0]$ and hence the system \eqref{eq:xtimepert}--\eqref{eq:momenttimepert} is a system of first-order, linear ordinary differential equations with constant coefficients for a fixed value of $S_0$. Therefore, we can integrate the system over the interval $[0, \widehat{L}_0]$. However, observe that only the boundary conditions for $x_\mathrm{dyn},\ y_\mathrm{dyn}$ and $\theta_\mathrm{dyn}$ are known; that is, they must vanish at the endpoints $S_0 = 0$ and $S_0 = \widehat{L}_0$. To overcome this, we use a method that has been used in previous buckling analyses \cite{BenAmar2005,moulton2011circumferential}. 

We seek a solution of the form $\mathbf{x}_\mathrm{dyn} = b_1\mathbf{x}_\mathrm{dyn}^{(1)} + b_2\mathbf{x}_\mathrm{dyn}^{(2)} + b_3\mathbf{x}_\mathrm{dyn}^{(3)}$, where, at $S_0 = 0$, $\mathbf{x}_\mathrm{dyn}^{(1)},\ \mathbf{x}_\mathrm{dyn}^{(2)}$ and $\mathbf{x}_\mathrm{dyn}^{(3)}$ satisfy
\begin{align}
\mathbf{x}_\mathrm{dyn}^{(1)}(0) = (0, 0, 0, 1, 0, 0)^T, \quad \mathbf{x}_\mathrm{dyn}^{(2)}(0)  = (0, 0, 0, 0, 1, 0)^T, \quad \mathbf{x}_\mathrm{dyn}^{(3)}(0)  = (0, 0, 0, 0, 0, 1)^T.\label{eq:xdyn123}
\end{align}
Therefore, the solution components $\mathbf{x}_\mathrm{dyn}^{(1)},\ \mathbf{x}_\mathrm{dyn}^{(2)}$ and $\mathbf{x}_\mathrm{dyn}^{(3)}$ are linearly independent and thus each solution may be solved for independently as functions of $\sigma^2$. To determine $b_1$, $b_2$ and $b_3$ and consequently our final solvability condition, we impose the boundary condition that $x_\mathrm{dyn},\ y_\mathrm{dyn}$ and $\theta_\mathrm{dyn}$ vanish at $S_0 = \widehat{L}_0$, leading to the matrix equation
\begin{align}
\left(\begin{array}{ccc}x_\mathrm{dyn}^{(1)} & x_\mathrm{dyn}^{(2)} & x_\mathrm{dyn}^{(3)} \\
				y_\mathrm{dyn}^{(1)} & y_\mathrm{dyn}^{(2)} & y_\mathrm{dyn}^{(3)} \\ 
				\theta_\mathrm{dyn}^{(1)} & \theta_\mathrm{dyn}^{(2)} & \theta_\mathrm{dyn}^{(3)}\end{array}\right)\left(\begin{array}{c}b_1\\b_2\\b_3\end{array}\right) = \left(\begin{array}{c}0\\0\\0\end{array}\right) \quad \mbox{ at }\quad S_0 = \widehat{L}_0.\label{eq:b1b2b3condition}
\end{align}
A solution to Equation \eqref{eq:b1b2b3condition} exists if and only if the determinant of the left hand side matrix is equal to zero for a given $\sigma^2$. Therefore, if there is a value of $\sigma^2 < 0$ such that the determinant vanishes, then the buckled solution $\mathbf{x}_\mathrm{eq}$ is unstable, as the time perturbation grows exponentially as $T \to \infty$.
\section{Energies}
\label{sec:energy}
Here, we give the forms of the energy functionals that are considered in Sections \ref{sec:parameterheterogeneity}. We then show how these defined energies change for heterogeneity in rod stiffness and growth, helping to elucidate the observed changes to rod shape in Figure \ref{fig:heterogeneitysummarised}. 

The total energy of the system, $\mathcal{E}$, comprises contributions from bending, stretching, and the underlying foundation. 

In the reference configuration, these dimensional individual energy densities may be written respectively as:
\begin{align}
\mathrm{U}^\mathrm{B} =\frac{EI}{2}\left(\frac{\partial\theta}{\partial S}\right)^{2}, \quad \mathrm{U}^\mathrm{S} =  \frac{EA}{2}(\alpha-1)^2, \quad \mathrm{U}^\mathrm{F} = \frac{Ek}{2\gamma}\left[\left(x-\frac{S}{\gamma}\right)^2 + y^2\right].\label{eq:energydensitiesrefconfig}
\end{align}
After nondimensionalisation, we write the energy densities as
\begin{align}
\mathrm{U}^\mathrm{B} &=\frac{\widehat{E}}{2}\left(\frac{\partial\theta}{\partial S}\right)^{2}, \quad \mathrm{U}^\mathrm{S} &=  \frac{\widehat{E}}{2}(\alpha-1)^2, \quad \mathrm{U}^\mathrm{F} &= \frac{\widehat{k}}{2\gamma}\left[\left(x-\frac{S}{\gamma}\right)^2 + y^2\right],\label{eq:nondimenergydensitiesrefconfig}
\end{align}
where $\widehat{E}$ and $\widehat{k}$ represent the dimensionless rod and foundation stiffness repectively (note that in the homogeneous case, $\widehat{E} =1$).
Therefore, the total energy density is given by
\begin{align}
\mathrm{U}^\mathrm{total} = \mathrm{U}^\mathrm{B} + \mathrm{U}^\mathrm{S} + \mathrm{U}^\mathrm{F} + F\left(\frac{\partial x}{\partial S} - \alpha\cos\theta\right) + G\left(\frac{\partial x}{\partial S} - \alpha\sin\theta\right).\label{eq:totalenergydensity}
\end{align}
The latter two terms correspond to the geometric constraints placed on the rod, with the horizontal and vertical forces $F$ and $G$ acting as Lagrange multipliers. However, these contributions will vanish after energy minimisation. Hence, the total energy $\mathcal{E}^\mathrm{total}$ is given by 
\begin{align}
\mathcal{E}^\mathrm{total} = \mathcal{E}^\mathrm{bend}+\mathcal{E}^\mathrm{stretch}+\mathcal{E}^\mathrm{foundation}, 
\end{align}
where
\begin{align}
 &\mathcal{E}^\mathrm{bend} = \int^{\widehat{L}_0}_0\mathrm{U}^\mathrm{B}\gamma dS_0,\\
 &\mathcal{E}^\mathrm{stretch} =  \int^{\widehat{L}_0}_0\mathrm{U}^\mathrm{S}\gamma dS_0,\\
 &\mathcal{E}^\mathrm{foundation} =  \int^{\widehat{L}_0}_0\mathrm{U}^\mathrm{F}\gamma dS_0.
\end{align}

\begin{figure}[t!]
\centering
\captionsetup{width=0.85\textwidth}
\includegraphics[width=0.7\textwidth]{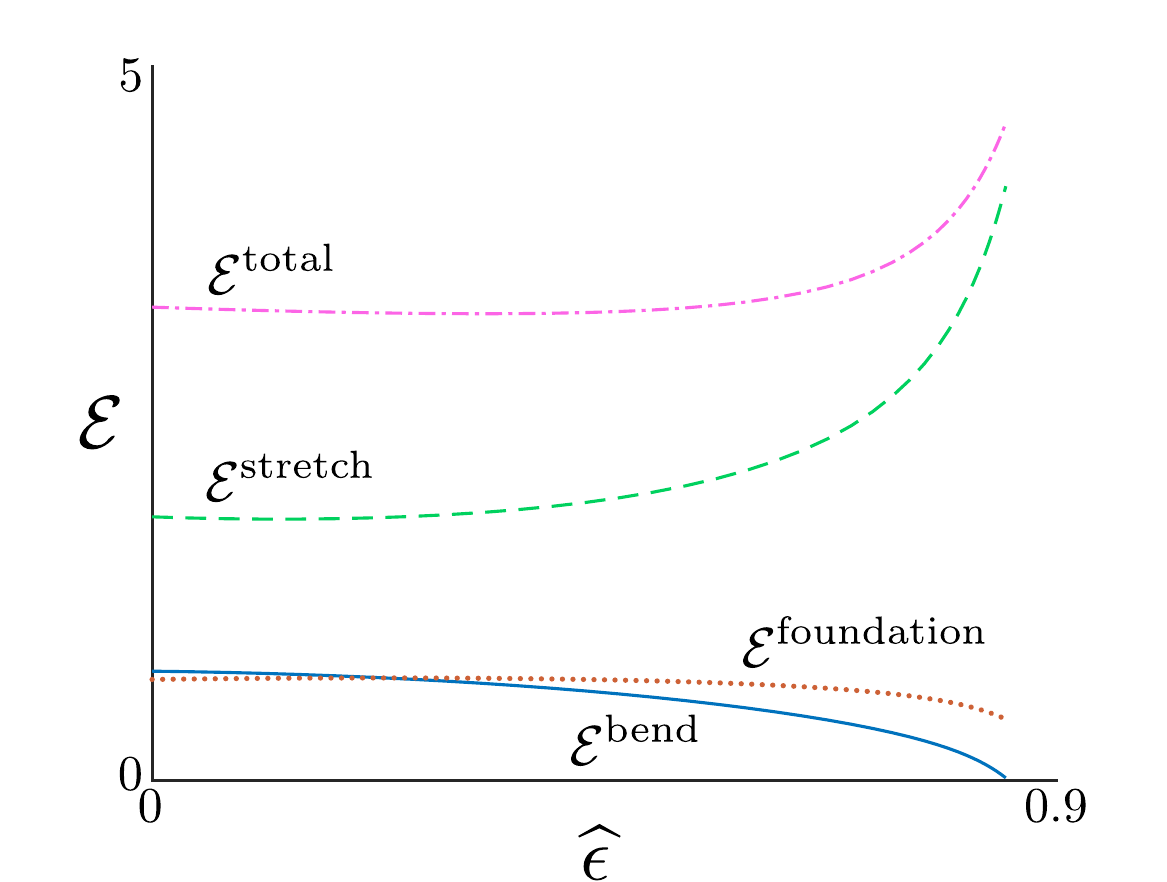}
\caption{\textbf{Energy plot for heterogeneous rod stiffness}. The bending energy $\mathcal{E}^\mathrm{bend}$ (blue, solid), stretch energy $\mathcal{E}^\mathrm{stretch}$ (green, dashed), $\mathcal{E}^\mathrm{foundation}$ (brown, dotted), and total energy $\mathcal{E}^\mathrm{total}$ (pink, dot-dashed) have been plotted. The parameters are the same as those in Figure \ref{fig:heterogeneitysummarised}. We see that $\mathcal{E}^\mathrm{stretch}$ increases while $\mathcal{E}^\mathrm{bend}$ and $\mathcal{E}^\mathrm{foundation}$ decrease for increasing heterogeneity.}
\label{fig:youngsheterogeneityenergyplot}
\end{figure}

\begin{figure}[t!]
\centering
\captionsetup{width=0.85\textwidth}
\includegraphics[width=0.65\textwidth]{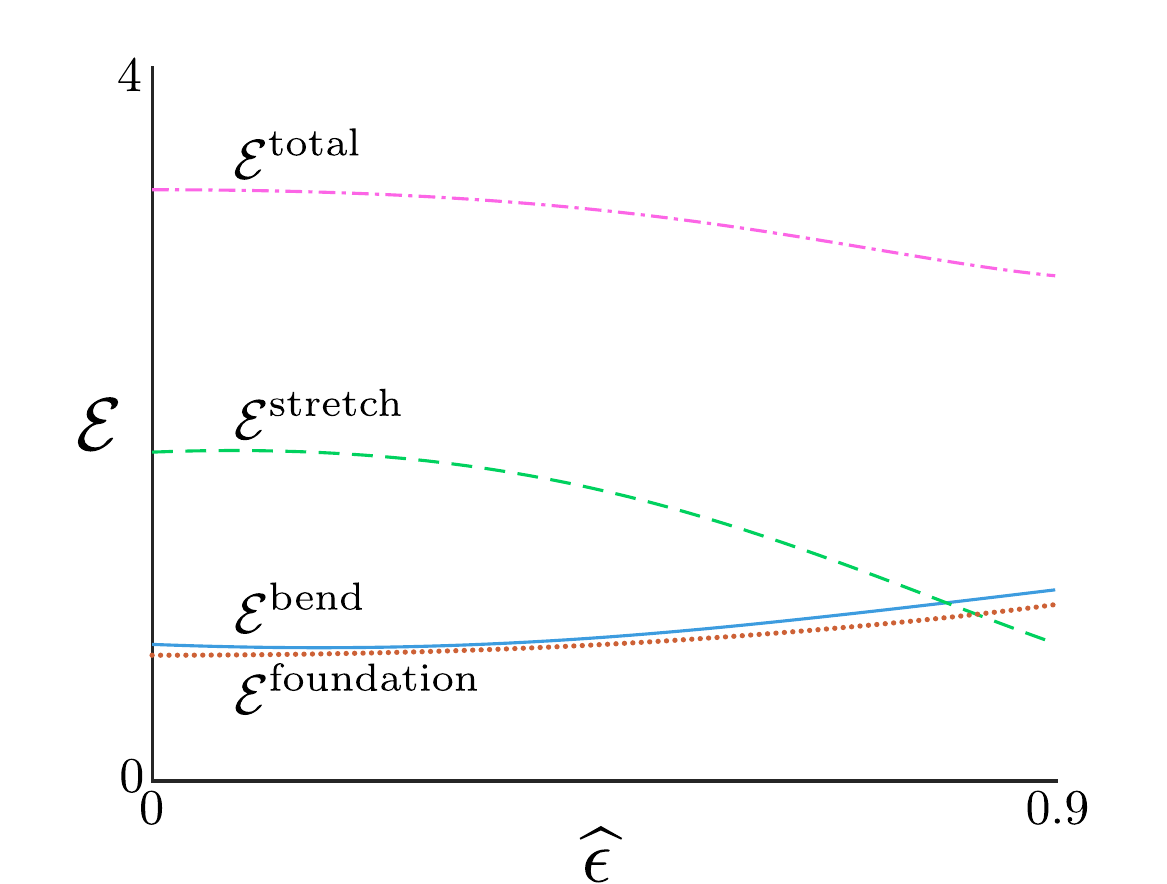}
\caption{\textbf{The effect of growth heterogeneity on energy for $\xi(S_0) = \cos\left(\frac{2\pi S_0}{\widehat{L}_0}\right)$}. The bending energy $\mathcal{E}^\mathrm{bend}$ (blue, solid), stretch energy $\mathcal{E}^\mathrm{stretch}$ (green, dashed), $\mathcal{E}^\mathrm{foundation}$ (brown, dotted), and total energy $\mathcal{E}^\mathrm{total}$ (pink, dot-dashed) have been plotted. The parameters are the same as those in Figure \ref{fig:heterogeneitysummarised}. As growth heterogeneity increases, $\mathcal{E}^\mathrm{stretch}$ decreases drastically, with $\mathcal{E}^\mathrm{bend}$ and $\mathcal{E}^\mathrm{foundation}$ increasing slightly.}
\label{fig:growthheterogeneityenergyplot}
\end{figure}

Figures \ref{fig:youngsheterogeneityenergyplot}--\ref{fig:growthheterogeneityenergyplot} illustrate how the different energetic contributions change for increasing heterogeneity amplitude $\widehat{\epsilon}$, in rod stiffness heterogeneity and growth heterogeneity, respectively. As $\widehat{\epsilon}$ increases, we observe tradeoffs between bending and foundation energy, and stretching energy. In Figure \ref{fig:youngsheterogeneityenergyplot}, the stretching energy increases, while bending and foundation energy decrease. This tradeoff helps to explain why compression dominates in the case of rod stiffness heterogeneity.
In the case of growth heterogeneity, the behaviour is qualitatively opposite, due to the transition to tension in the middle region of the rod. 


\dem{\section{The inverse problem}
\label{sec:inverseproblem}}
\dem{In this section, we describe our approach to the inverse problem discussed briefly in Section \ref{sec:discussion}, and the method of producing Figure \ref{fig:heterogeneitysignatures}. We chose a candidate shape with embedded heterogeneity, in this case, foundation heterogeneity following Figure \ref{fig:heterogeneitysummarised}(b), and we tried to match that shape by varying the heterogeneity in the rod stiffness and growth, in the form of \eqref{eq:arbitraryheterogeneity}, as well as the net growth $\gamma_0$, based on the results of Section \ref{sec:parameterheterogeneity}. The baseline foundation stiffness $\widehat{k}_0$ and rod length $\widehat{L}_0$ were set to $\widehat{k}_0 = 0.04$ and $\widehat{L}_0 = 20$, respectively.} 

\dem{Figure \ref{fig:heterogeneitysummarised}(c) showed that when net growth is unchanged, growth heterogeneity redistributes the material of the rod away from regions of low growth to those with high growth. This suggests that in order to match the candidate shape, growth must be high in the middle region and low in the outer regions. For rod stiffness heterogeneity, we wanted to localise the buckling to regions of softened rod stiffness, but to counteract the enhanced compressed seen in Figure \ref{fig:heterogeneitysummarised}(b), we increased the net growth. }

\dem{Here we have only explored this problem in a preliminary manner. An extensive search of the heterogeneity parameter space is intrinsically challenging and time-consuming, as the heterogeneity can in principle take any form, and the morphology produced for a given form can only be obtained as the result of numerical path continuation from a known solution. While a more detailed investigation would be a worthwhile focus for another paper, here we have chosen heterogeneities in the {\it ad hoc}, but intuitively guided, manner outlined above. Our prescribed heterogeneity choices were: for foundation stiffness heterogeneity, we set $\gamma = 1.9$, $\widehat{\epsilon} = 0.75$ and $\xi(S_0) = \cos(2\pi S_0/\widehat{L}_0))$; for rod stiffness heterogeneity, $\gamma$ was increased to $\gamma = 2.45$, $\widehat{\epsilon} = 0.9$, and $\xi(S_0) = 1 - \exp\left(-16(S_0 - 0.5\widehat{L}_0)^2/\widehat{L}_0^2\right)$; and for growth heterogeneity, we prescribed $\gamma_0 = 1.9$, $\widehat{\epsilon} = 0.9$, and $\xi(S_0) = -\cos(2\pi S_0/\widehat{L}_0))$.}

\dem{Once we had matched the shapes to a sufficient degree, we examined all other features, such as the bending moment $m$, force components $F$, $G$, etc, seeking distinguishing characteristics. From this, we identified two salient features: the axial stress $n_3$ in the outer region, where there was minimal growth (Fig. \ref{fig:heterogeneitysignatures}(b)); and the foundation energy density $\mathrm{U}^\mathrm{F}$ in the middle section, where the foundation had been softened locally (Fig. \ref{fig:heterogeneitysignatures}(c)).}

\end{document}